\documentclass{aa}

\usepackage{natbib}
\bibpunct{(}{)}{;}{a}{}{,}

\usepackage{graphicx}
\usepackage{booktabs}
\usepackage{txfonts}
\usepackage{orcidlink}
\usepackage[caption=false]{subfig}
\usepackage{float}
\usepackage{threeparttable}
\usepackage[utf8]{inputenc}
\usepackage{xspace}
\DeclareUnicodeCharacter{03B1}{\ensuremath{\alpha}}

\usepackage{hyperref}
\hypersetup{hidelinks}

\newcommand{\lya}{Ly$\alpha$\xspace}
\newcommand{\pixcgs}{erg~s$^{-1}$~cm$^{-2}$~\AA$^{-1}$~pix$^{-1}$}
\newcommand{\sblcgs}{erg~s$^{-1}$~cm$^{-2}$~arcsec$^{-2}$}
\newcommand{\kms}{km~s$^{-1}$\xspace}
\newcommand{\SB}{\mathrm{SB}}

\begin{document} 

\title{Cosmic web Ly$\alpha$ emission in a sample of overdense regions}
\titlerunning{Emission from LAE overdensities}

\authorrunning{Tornotti et al.}

    \author{Davide Tornotti$\,$\orcidlink{0009-0001-3388-8742}
        \inst{1}\thanks{\href{mailto:d.tornotti@campus.unimib.it}{d.tornotti@campus.unimib.it}},
    Michele Fumagalli$\,$\orcidlink{0000-0001-6676-3842} \inst{1,3},
    Matteo Fossati$\,$\orcidlink{0000-0002-9043-8764} \inst{1,2},  
    Roland Bacon$\,$\orcidlink{0000-0001-6148-0915} \inst{4},
    Fabrizio Arrigoni Battaia$\,$\orcidlink{0000-0002-4770-6137} \inst{5},
    Rajeshwari Dutta$\,$\orcidlink{0000-0002-6095-7627} \inst{6},
    Michele Fulghieri$\,$\orcidlink{0009-0000-0313-4459} \inst{1},
    C\'eline P\'eroux$\,$\orcidlink{0000-0002-4288-599X} \inst{7,8},
    Marc Rafelski$\,$\orcidlink{0000-0002-9946-4731} \inst{9,10},
    Mitchell Revalski$\,$\orcidlink{0000-0002-4917-7873} \inst{11}
    }

\institute{Dipartimento di Fisica ``G. Occhialini'', Universit\`a degli Studi di Milano-Bicocca, Piazza della Scienza 3, 20126 Milano, Italy \label{unimib}
\and
    INAF - Osservatorio Astronomico di Brera, Via Brera 28, I-20121 Milano, Italy
\and
    INAF - Osservatorio Astronomico di Trieste, Via G. B. Tiepolo 11, I-34143 Trieste, Italy
\and 
    Univ. Lyon, Univ Lyon1, ENS de Lyon, CNRS, Centre de Recherche Astrophysique de Lyon UMR5574, 9 avenue Charles Andr{\'e}, 69230 Saint-Genis-Laval, France
\and 
    Max-Planck-Institut f\"ur Astrophysik, Karl-Schwarzschild-Str. 1,
    D-85748 Garching bei M\"unchen, Germany
\and
    IUCAA, Postbag 4, Pune 411007, Ganeshkind, India
\and 
    European Southern Observatory, Karl-Schwarzschild-Str. 2, D-85748 Garching bei M\"unchen, Germany
\and 
    Aix Marseille Universit{\'e}, CNRS, LAM (Laboratoire d'Astrophysique de Marseille) UMR 7326, F-13388 Marseille, France
\and 
    Space Telescope Science Institute, 3700 San Martin Drive, Baltimore, MD 21218, USA
\and
    Department of Physics and Astronomy, Johns Hopkins University, Baltimore, MD 21218, USA
   }

\abstract{
We present a complete and homogeneous analysis of the \lya emission properties of the cosmic web at $3 \lesssim z \lesssim 5$ as a function of the overdensity of Ly$\alpha$ emitters (LAEs) in the MUSE Ultra Deep Field (MUDF) and the MUSE Extremely Deep Field (MXDF). We identified 41 overdensities, probing environments that are $\delta \approx 2-10$ times denser than the field. We searched for extended emission down to surface brightness (SB) levels of $3-5\times 10^{-20}$~\sblcgs, revealing filamentary structures beyond the scale of the circumgalactic medium, 
confirming that LAEs act as signposts of the cosmic web. A trend with overdensity emerges: non-detections are mainly at $\delta < 2$, a $\approx 1$~dex scatter in SB appears for $2 \lesssim \delta \lesssim 3.5$, and detections reach a maximum intrinsic SB of $\approx 2\times 10^{-19}$~\sblcgs\ for $\delta \gtrsim 3.5$. Emitting regions occupy a fraction of projected area $f_{s}\approx 0.20-0.25$ inside filaments, yielding a cosmological incidence for \lya emission in the cosmic web $\ell(\rm Ly\alpha)$ of $\approx 1.22$, which is similar to the one of partially neutral Lyman limit systems. This analysis indicates that the emitting gas is partially ionized at moderate densities ($n_H \approx 10^{-3} - 10^{-1} \text{ cm}^{-3}$), likely tracing the denser spines of intergalactic filaments and embedded substructures. Finally, we forecast how oriented stacking in larger samples of overdensities from shallower observations could yield an expanded view of the cosmic web before the next-generation wide-field spectroscopic instruments become operational.}


\maketitle
\section{Introduction}

Within the current $\Lambda$ cold dark matter ($\Lambda$CDM) structure formation framework, galaxies are understood as open and dynamically evolving systems whose growth is regulated by the continuous interplay with their large-scale environment. Rather than evolving in isolation, galaxies reside at the nodes of a vast and interconnected network of filaments -- the cosmic web \citep{bond_how_1996}. 
These filaments host a significant fraction of the baryons in the Universe and provide the supply through which pristine gas is accreted from the intergalactic medium (IGM) into the galactic potential wells \citep{keres_how_2005, dekel_cold_2009}. As a consequence, key physical properties of galaxies (e.g., stellar mass and star formation rate) are the integrated result of a complex baryonic cycle that extends far beyond the stellar disk, involving outflows and wind recycling in the circumgalactic medium (CGM) and inflow from the surrounding IGM \citep[e.g.,][]{Dave2012, Lilly2013, tumlinson_circumgalactic_2017}. In this context, characterizing the spatial distribution of gas on scales beyond individual virial radii and tracing the filamentary structures connecting galaxies is essential to understanding how the cosmic web regulates galaxy assembly and evolution.

Despite its fundamental role in galaxy evolution, directly detecting the diffuse gas associated with the cosmic web in emission remains one of the most significant observational challenges. This is primarily due to the low gas densities expected in the IGM, where the hydrogen density is only mildly overdense, typically a factor of $1-10$ above the cosmic mean \citep[e.g.,][]{Hernquist1996, Miralda-Escude1996}. 
Historically, our understanding of the large-scale gas distribution has been built primarily through absorption-line studies along quasar sight lines \citep[e.g.,][]{hennawi_quasars_2013, Rudie2019, peroux_cosmic_2020}. 
In particular, observations of the \lya forest have enabled statistical reconstructions of the IGM down to very low neutral hydrogen column densities \citep[$N_\mathrm{HI}\approx 10^{14-15} \rm ~cm^{-2}$, e.g.,][]{rauch_lyman_1998,  mcquinn_evolution_2016}. However, absorption studies probe the cosmic web gas only along sparse one-dimensional lines of sight and lack the spatial information required to resolve the geometry of the gas on sub-megaparsec scales. 

Emission studies offer a complementary perspective by directly mapping the spatial distribution of the diffuse gas. Theoretical and numerical works predict that filaments emit \lya radiation through a combination of physical mechanisms, including recombination following photoionization from the ultraviolet background (UVB), photoionization by local sources, and collisional excitations of cooling gas \citep[e.g.,][]{Kollmeier2010, Elias2020, Byrohl2023}. In the absence of strong local ionizing sources, UVB-induced fluorescence is expected to produce \lya surface brightness (SB) levels of $\approx 10^{-20}$ \sblcgs\ in dense gas \citep[e.g.,][]{Hogan1987, Gould1996} at redshift $z\approx 3$, accounting for the $(1+z)^4$ cosmological SB dimming \citep{Tolman1930}.

Observationally, narrow-band (NB) imaging surveys have been used to search for extended \lya emission \citep[e.g.,][]{cantalupo_cosmic_2014} over wide fields of view, providing constraints on the incidence of emitting structures. 
A major observational breakthrough has been enabled by the advent of a new generation of sensitive integral field unit (IFU) spectrographs, such as the Multi Unit Spectroscopic Explorer \citep[MUSE,][]{Bacon2010} on the Very Large Telescope (VLT) and the Keck Cosmic Web Imager \citep[KCWI,][]{Morrissey2012}. By providing the simultaneous acquisition of spatial and spectral information over a field of view (FoV) of order arcminutes, IFU observations allow the reconstruction of the full three-dimensional (position-position-velocity) distribution of the emitting gas.
Most of the observational efforts focused on extreme environments dominated by luminous quasars ($L_\mathrm{bol}\gtrsim 10^{45} \rm \, erg\,s^{-1}$) and massive protocluster regions, where the \lya signal is likely enhanced. These studies revealed the presence of giant \lya nebulae surrounding the quasars, with typical spatial extents of $\approx 100$ kpc, in some cases exhibiting filamentary morphologies, and characteristic SB levels of $\approx 10^{-17}-10^{-18}$ \sblcgs\ \citep[e.g.,][]{Borisova2016, Cai2017, cai_keckpalomar_2018, martin2019, ArrigoniBattaia2019, Fossati2021, Mackenzie2021, Gonzalez2025}. 
Such systems trace highly biased regions of the Universe, where a luminous active galactic nucleus (AGN) dominates the radiation field, enhancing the \lya emission of the surrounding CGM and facilitating its detection out to, and in some cases beyond, the virial radius at SB levels $\gtrsim 10^{-18}$ \sblcgs.

Particularly compelling evidence for the filamentary nature of diffuse \lya emission has been provided by observations of physically associated quasar pairs. In these systems, extended \lya emission is preferentially aligned with the axis connecting the two quasars, consistent with gas distributed along large-scale filaments of the cosmic web traced by the quasar pair \citep[e.g.,][]{Lusso2019, arrigoni_battaia_discovery_2019, herwig_qso_2024}. Pushing to lower SB limits ($\approx 3-5\times 10^{-20}$ \sblcgs), ultra-deep ($\gtrsim 140$ hr on source) MUSE observations of the MUSE Ultra Deep Field \citep[MUDF,][]{Fossati2019} have directly revealed a $\approx 700$ kpc cosmic web filament in emission connecting a quasar pair at $z\approx 3.22$ \citep{Tornotti2025c}. The emission, detected down to a mean SB of $8\times10^{-20}$ \sblcgs, has enabled a detailed characterization of the SB profiles and morphology of the intergalactic gas, including the transition radius from the CGM and IGM and the filament's width. 
Similarly, extended and filamentary Ly$\alpha$ structures tracing the megaparsec large-scale structure have been uncovered in highly overdense environments such as the SSA22 protocluster region at $z\approx3.1$, down to a SB of $3\times10^{-19}$ \sblcgs\  \citep{Umehata2019}. 

To obtain a representative view of the baryonic content in the large-scale filaments, it is necessary to move beyond rare and extreme environments and probe more typical regions of the Universe, where no evident luminous AGN or massive proto-cluster is present. In this regime, groups and overdensities of Ly$\alpha$ emitters (LAEs) can be used as tracers of the nodes of cosmic web filaments \citep[e.g.,][]{Matsuda2005, Soo2014, Im2024, Ramakrishnan2024}.
Recent findings have shown that diffuse \lya emitting gas can also be detected in these environments, in some cases revealing extended and filamentary structures aligned with LAE overdensities \citep[][]{Bacon2021, banerjee_2024, Tornotti2025b}.
However, the expected SB of the emission in such environments drops to $\rm SB \lesssim 10^{-20}-10^{-19}$ \sblcgs, pushing current instrumentation to its limits and requiring ultra-deep observations \citep[e.g.,][]{Bacon2021}. 
The challenge therefore lies in transitioning from targeted studies of rare, highly biased environments to a systematic characterization of more typical regions of the high-redshift Universe. 

By identifying groups and overdensities of LAEs, it becomes possible to perform a census of the diffuse gas surrounding these galaxies. This approach enables the quantification of the incidence, spatial extent, and morphology of \lya emission extending beyond individual virial radii, and allows us to assess whether such emission preferentially connects galaxies in a filamentary configuration. Importantly, both detections and non-detections provide constraints on the physical conditions of the CGM-IGM interface, offering key observational benchmarks for cosmological simulations of gas accretion and baryon cycling.

In this paper, we homogeneously combine the ultra-deep integral field spectroscopy (IFS) observations of the MUDF and of the MUSE Extremely Deep Field \citep[MXDF;][]{Bacon2021, Bacon2023} to investigate the distribution and properties of diffuse \lya emission associated with LAE overdensities in the redshift range $2.8 \lesssim z \lesssim 5$. By applying a consistent analysis framework to both datasets, we construct a census of extended \lya emission in typical galaxy environments, enabling a systematic exploration of the cosmic web in less extreme overdensities.

The paper is structured as follows. In Sect.~\ref{sect:obs-cat}, we describe the MUSE observations and the LAE catalogs. Sect.~\ref{sect:over-def} outlines the identification of LAE overdensities, while Sect.~\ref{sect:ext-emis} presents the extraction of the associated extended \lya emission. In Sect.~\ref{sect:analysis}, we describe the analysis of the extraction, and in Sect.~\ref{sect:discussion}, we discuss the results. Our conclusions are summarized in Sect.~\ref{sect:conclusions}.
Throughout this paper, we adopt a standard $\Lambda\mathrm{CDM}$ cosmology with $\Omega_m = 0.31$, and $H_0 = 67.7 \,\mathrm{km\,s^{-1}\,Mpc^{-1}}$ \citep{Planck2020}.

\section{Observations and LAE catalog}\label{sect:obs-cat}

The MUDF ($\text{RA} = 21^\text{h}:42^\text{m}\,24^\text{s}, \, \text{Dec} = -44^\circ:19':48''$) is a $142$-hour VLT/MUSE large program covering an area of approximately $2.1\times1.9\,\text{arcmin}^2$ in a target field with a pair of two bright quasars at $z\approx 3.22$. The final datacube reaches an observational depth of $\gtrsim 90$ hr (up to $\approx 120$ hr) in the central $\approx 0.55\,\text{arcmin}^2$ and a voxel root mean square of $3\times10^{-21}$ \pixcgs at $\approx 5200$ \AA. The observation strategy and data reduction process are detailed in \cite{Lusso2019, Fossati2021, Tornotti2025c}. In this work, we adopt the final datacube presented in \cite{Tornotti2025c}, in which both continuum sources and the $z\approx 3.22$ quasar point-spread functions have been removed with the techniques presented in, for example, \citet{Borisova2016} and \citet{ArrigoniBattaia2019}. 

\begin{figure}[h]
\centering
\resizebox{0.9\hsize}{!}{\includegraphics{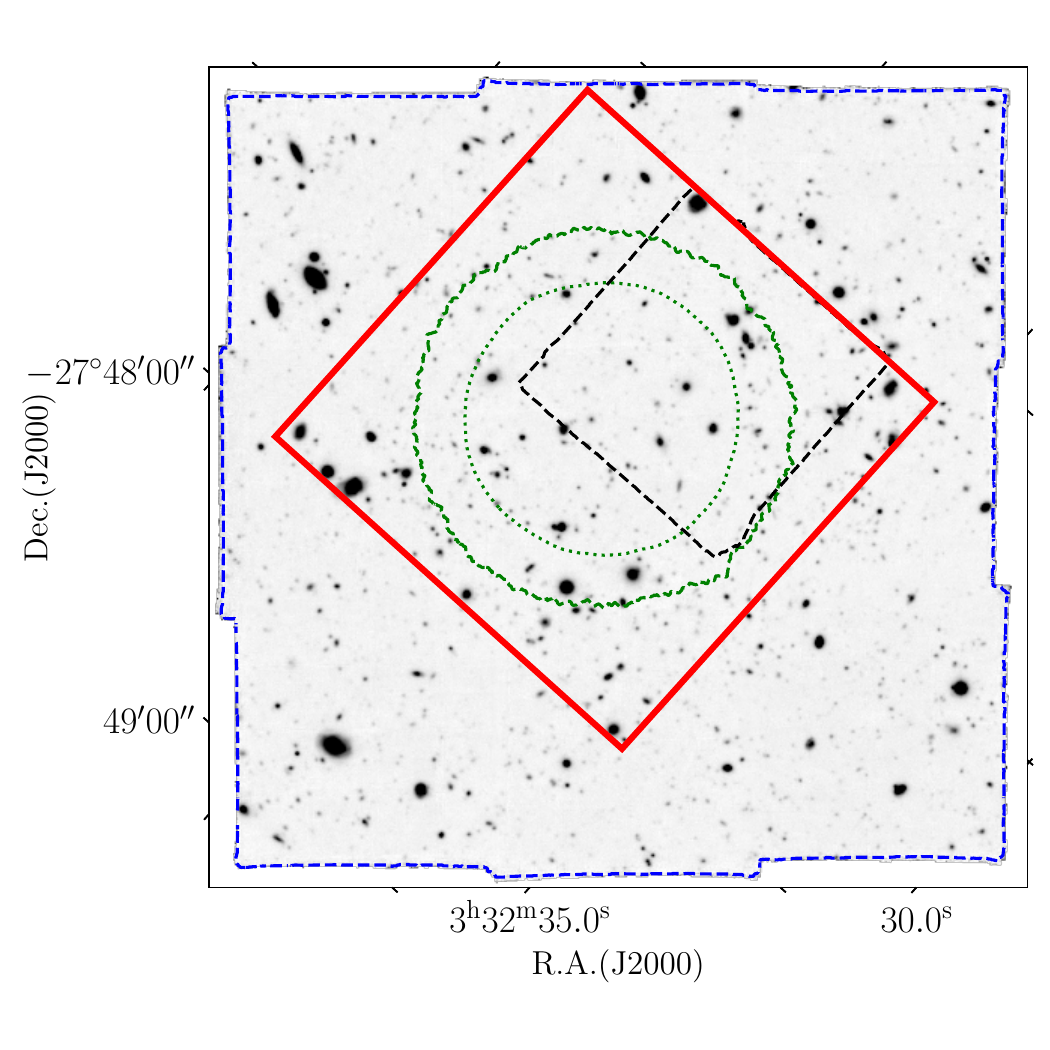}}
\vspace{-0.5 cm}
\caption{Survey footprints for the MOSAIC (dashed blue contour), UDF-10 (dashed black), and MXDF (dashed green) MUSE programs. The dotted green contour highlights the deepest central region of the MXDF ($\gtrsim 90$ hr, reaching $\approx 140$ hr). The solid red contour defines the MUDF-equivalent area selected for a consistent identification of LAE overdensities across the two datasets.}
\label{fig:mxdf-footprint}
\end{figure}

The MXDF is a $155$-hour VLT/MUSE program centered in $\text{RA} = 03^\text{h}:32^\text{m}\,39^\text{s}, \, \text{Dec} = -27^\circ:47':05''$ within the Hubble Ultra Deep Field (HUDF). It covers an approximately circular area with a radius of $\approx 40''$, reaching an observational depth $\gtrsim 90$ hr (up to $140$~hr) within the central $\approx 30 ''$. For details on observations and data reduction, see \cite{Bacon2021, Bacon2023}.
To perform a robust and homogeneous comparison between the MUDF and MXDF datasets, we address the potential difficulty of detecting overdensities in a small footprint. We extend the MXDF footprint by incorporating the overlapping and adjacent UDF-10 ($30$-hr depth) and MOSAIC ($10$-hr depth) observations \citep{Bacon2017, Bacon2023}, maximizing the observational depth of the covered area. This allows us to define a search area of $\approx 3.2 \rm\, arcmin^2$ (i.e., similar to the MUDF footprint with an exposure $>3$ hr) also for the HUDF region (bound by the red square in Fig. \ref{fig:mxdf-footprint}).
Indeed, while completeness differences -- which determine the number of detected LAEs -- are accounted for through selection function corrections, the physical interpretation of an overdensity depends on the area over which it is defined. A density fluctuation that appears significant over a larger footprint may lose its statistical standing when measured on a smaller scale, and vice versa. This standardized area ensures that our census of LAE overdensities remains consistent and comparable across both ultra-deep datasets.

\begin{figure*}
\sidecaption
\includegraphics[width=12cm]{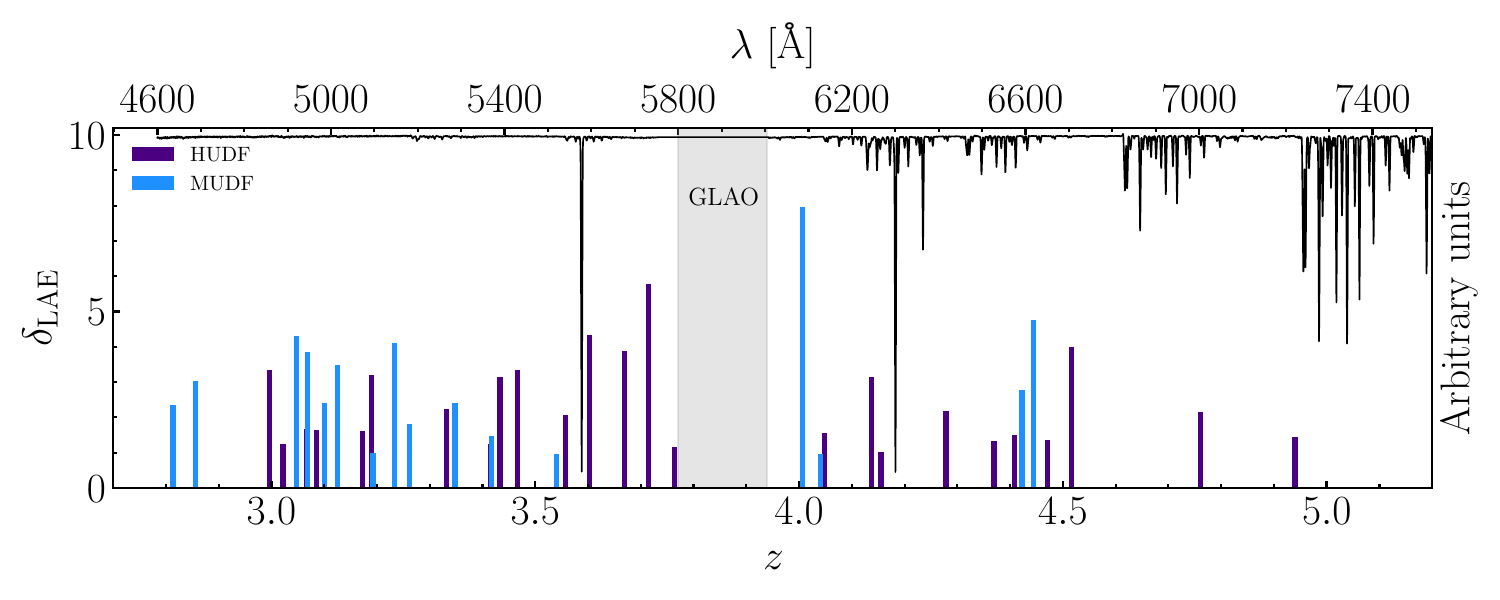}
\caption{Redshift distribution of all the identified LAE groups in the MUDF (blue) and MXDF (purple), with at least four members and $\delta\geq 1$. The bins are centered on the mean redshift of each group. A normalized sky spectrum (shown upside down) is shown on the top wavelength axis to highlight redshift regions affected by sky line residuals. 
The gray shaded area represents the redshift range excluded due to the Ground Layer Adaptive Optics (GLAO) module, which uses an artificial laser
guide star to improve image quality during the observations. Therefore, in this redshift interval, no data are available.}
\label{fig:z-distr}
\end{figure*}

The construction of the MUDF LAE catalog is described in detail in \cite{Tornotti2025a}. Briefly, we extract the line emitter candidates from the final continuum-subtracted datacube by identifying compact emission using the Spectral Highlighting and Identification of Emission (SHINE) algorithm \citep{Fossati2025}. The full set of extraction parameters is provided in \cite{Tornotti2025a}. We then classify all candidates based on their integrated signal-to-noise ratio ($ISN$), retaining only those with $ISN \geq 7$ \citep[following][]{Lofthouse2020}. Each candidate exceeding this threshold is then visually inspected independently by three authors (DT, MF, MFo). By distinguishing between emitters with line profiles compatible with Ly$\alpha$ emission and those that match lower redshift transitions observed by MUSE (e.g., [\ion{O}{II}], \ion{C}{III}], [\ion{O}{III}], \ion{H}{$\beta$}), we both identify reliable LAE and keep track of all the others line emitters. After this validation process, a sample of $212$ sources are spectroscopically confirmed to be LAEs within the MUSE-probed redshift range $2.8 \lesssim z \lesssim 6.6$.
While a comprehensive LAE catalog for the HUDF region is publicly available and described in \cite{Bacon2023}, we perform a re-extraction to ensure consistency with the MUDF analysis. We extend the SHINE-based extraction procedure -- originally applied to the continuum-subtracted MXDF cube in \cite{Tornotti2025a} -- to the UDF-10 and MOSAIC continuum-subtracted cubes. This analysis is restricted to the MUDF-like area defined in Fig. \ref{fig:mxdf-footprint} and follows the exact same workflow used for the MUDF, including the $ISN$ definition and minimum threshold ($ISN>7$). By cross-matching our identified sources with the \cite{Bacon2023} catalogs, we obtained a final sample of $507$ LAEs in the selected HUDF region within the redshift range $2.8 \lesssim z \lesssim 6.6$. 
We recover $\approx 70\%$ of the sources from the \cite{Bacon2021} catalog within our analyzed area. This difference is a direct consequence of our more conservative approach ($ISN \geq 7$) which prioritizes sample purity, with the vast majority of the unrecovered sources belonging to their lowest-confidence class.

\section{Overdensities of LAEs}\label{sect:over-def}

The LAEs are not distributed uniformly; rather, they show clustering both in spatial and in redshift space, giving rise to overdensities \citep[e.g.,][]{Matsuda2005, Soo2014, Bielby2016, Shi2019, Ouchi2020, Herrera2025}. In this work, we aim to verify to what extent these overdensities are signposts of filaments. 
A first step is therefore to search for groups of LAEs at $2.8 \lesssim z \lesssim 5 $, where the availability of a robust luminosity function down to luminosities as low as $\log(L_\mathrm{Ly\alpha}/\rm erg\,s^{-1})\approx 41$ ensures a reliable characterization of the underlying density field \citep{Tornotti2025a}, also avoiding multiple skylines (see Fig. \ref{fig:z-distr}).  We use a friends-of-friends approach \citep{Huchra&Geller1982}, which considers a galaxy as a member of a group based on linking lengths in both projected radial distance, $\Delta R$, and redshift space velocity, $\Delta v$. We adopt $\Delta R=500$ kpc and $\Delta v=400$ \kms to account for both the area covered by the observations ($\gtrsim 1\,\rm arcmin^2$) and the expected diameter ($\gtrsim 340$ kpc) and one-dimensional velocity dispersion ($\gtrsim 360$ \kms) of dark matter halos with masses up to $10^{13}-10^{13.5}\,M_\odot$, which typically host rich overdensities of galaxies at $z\gtrsim3$ \citep[e.g.,][]{Chiang2013, Overzier2016}. With these parameters, we obtain $49$ groups with four or more associated galaxies, $16$ from the MUDF and $33$ from the HUDF selected area. All the groups we identify in the HUDF are included in \cite{Bacon2021}, except for an additional seven groups (see Table \ref{tab:combined_overdensities}). This is likely due to the combined effect of using a different identification approach and of searching the full HUDF area in \cite{Bacon2021}. 
We also tested the classification of the groups by varying the two linking lengths within $\pm 100$ kpc and $\pm 100$ \kms and found no strong variation in the resulting group definitions. 
Furthermore, we verified that adopting the systemic redshifts of LAEs -- estimating using the empirical correlations from \cite{Verhamme2018} -- yields highly consistent groups. Since the systemic correction applies a similar shift to the bulk of the population (with a scatter $\approx 100$ \kms, well within our $\Delta v = 400$ \kms linking length), the relative velocities remains largely unchanged, confirming the robustness of grouping based on \lya redshifts \citep[see also the discussion in][]{Bacon2021}. A key quantity we aim to estimate for the identified groups is the associated overdensity, $\delta_\mathrm{LAE}$, which is defined by $\delta_\mathrm{LAE}=N^\mathrm{obs}_\mathrm{LAE}/\bar{N}_\mathrm{LAE}-1$, where $N^\mathrm{obs}_\mathrm{LAE}$ is the observed number of galaxies within a given cosmological volume and $\bar{N}_\mathrm{LAE}$ is the expected number of galaxies in the same volume. To compute $\bar{N}_\mathrm{LAE}$, we use the LAE luminosity function in \cite{Tornotti2025a}, derived by combining multiple MUSE datasets, including the MUSE ultra-deep fields and probing the faint end down to $\log(L/\rm erg\,s^{-1})\approx 41$. Specifically, we use the median luminosity functions obtained in the distinct redshift intervals $3\leq z \leq 4$ and $4<z \leq 5$. We then consider the corresponding selection functions that account for the varying sensitivity and depth of the observed area. While the selection functions for the MUDF and MXDF footprints were previously computed \citep{Tornotti2025a}, the current analysis includes the UDF-10 and a subregion of the MOSAIC, as shown in Fig. \ref{fig:mxdf-footprint}. Consequently, we have consistently derived the new selection functions for these specific areas with the same methodology (see Fig. \ref{fig:selfunc}).

Finally, we calculate $\bar{N}_\mathrm{LAE}$ within a fixed volume defined by the selected area and a redshift interval corresponding to a velocity window of $1000$ \kms, centered on the mean redshift of each group. This corresponds to $\approx 12.3$ cMpc at redshift $\approx3.5$. The adopted velocity window provides an optimal balance, ensuring that the full clustering signal of the observed LAE groups is captured, without excessive smoothing. We then consider as significant overdensities those with $\delta \geq 1$, obtaining the final sample of $41$ overdensities. In Fig. \ref{fig:z-distr} we show the redshift distribution of the identified overdensities and in Table \ref{tab:combined_overdensities} we summarize their main properties. 

\begin{figure}[t]
\centering
\vspace{-0.4cm}
\resizebox{0.9\hsize}{!}{\includegraphics{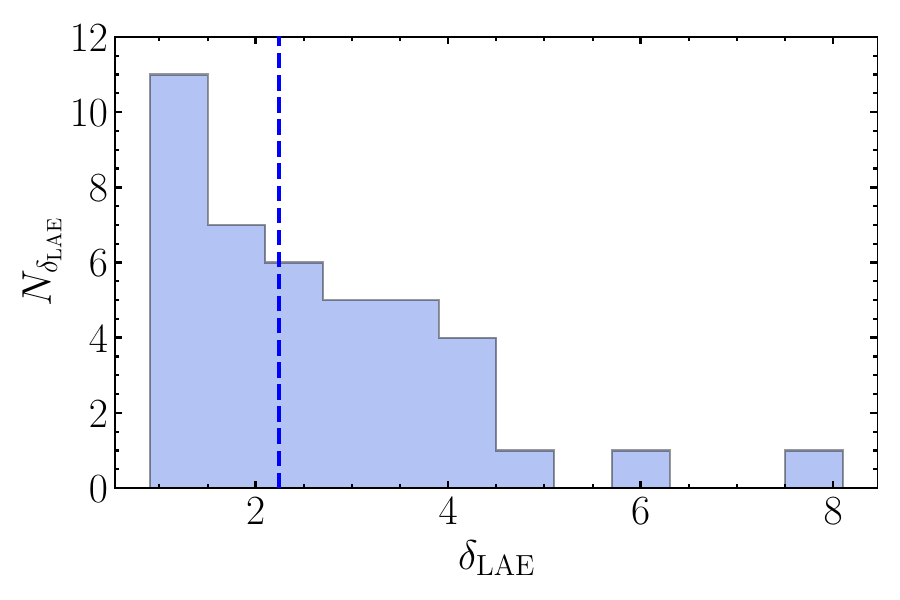}}
\vspace{-0.4cm}
\caption{Distribution of $\delta_\mathrm{LAE}$ obtained for all the $41$ identified overdensities. The dashed blue vertical line indicates the median of the distribution.}
\label{fig:delta-distr}
\vspace{-0.4 cm}
\end{figure}

As illustrated in Fig.~\ref{fig:delta-distr}, our combined sample probes a diverse range of cosmic environments, spanning regions that are $\approx 2$ to $10$ times denser than the average field, with a median value of $\approx 2.2$. 
We note that the absolute value of the overdensity at $z\approx 4$ analyzed in \cite{Tornotti2025b} is about a factor of $\approx 3$ lower here, corresponding to a $\approx 2 \sigma$ deviation from the previously reported value. This difference arises from a more refined approach adopted in the present work, which leads to a higher estimate of $\bar{N}_\mathrm{LAE}$ and, consequently, a lower $\delta_\mathrm{LAE}$. In particular, (i) the volume used to compute $\bar{N}_\mathrm{LAE}$ is slightly larger due to a larger velocity window (increased from $900$ \kms to $1000$ \kms), and (ii) we now rely on the LAE luminosity function from \cite{Tornotti2025a} combined with spatially varying selection functions that accurately capture the depth variations across the MUDF field -- especially within the ultra-deep $\approx 0.5~\mathrm{arcmin}^2$ region, where the luminosity function is integrated down to $\approx 10^{41}\, \mathrm{erg\,s^{-1}}$ rather than adopting a single averaged $50\%$ completeness limit at $10^{41.5} \, \mathrm{erg\,s^{-1}}$. These refinements result in a more accurate estimate of the expected number of LAEs within each region. Overall, while the absolute normalization of $\delta_\mathrm{LAE}$ is sensitive to the specific assumptions adopted in the computation, the relative comparison across the sample is the salient characteristic of our analysis. Because all overdensities are consistently derived using the same methodology, the relative differences between the groups are robust. 

\section{Extended Ly$\alpha$ emission} \label{sect:ext-emis}
The assumption we test in this work is that the identified overdensities are signposts for cosmic web filaments. We therefore aim to investigate the possible presence of extended \lya emission down to SB levels of $3-5\times10^{-20}$ \sblcgs. These limits are achievable within the ultra-deep region ($>90 \rm \, hr$) of the MUDF and MXDF. While such deep investigation have been previously conducted in the MXDF region \citep[e.g.,][]{Bacon2021}, we now include the MUDF sample and perform a systematic extraction across both fields to ensure a fully homogeneous procedure for each overdensity. We follow a method similar to that described in \cite{Tornotti2025b}, with additional robustness tests introduced to validate the extractions, given the lower-density nature of some overdensities. 

Based on the final continuum-subtracted datacubes we apply the SHINE algorithm tuned for extended emission extraction \citep[see][for further details]{Tornotti2025b}. For each overdensity, we extract an initial slice of the cube encompassing $88$ to $92$ spectral channels (or $110$ to $115$ \AA) centered on its mean redshift. This wide spectral slice provides a sufficient padding to avoid edge effects during the three-dimensional voxel grouping. We identify connected voxels ($> 4000$) with a signal-to-noise ratio ($S/N$) greater than $2$, as usually done in the literature \citep[e.g.,][]{Borisova2016, ArrigoniBattaia2019}, requiring at least $2000$ minimum number of spatial pixels (corresponding to a minimum area of $80 \rm \, arcsec^2$ or $\approx 5\times10^3 \, \mathrm{kpc}^2$ at $z=3$). To enhance the sensitivity to low-SB emission, we apply a two-dimensional Gaussian spatial smoothing kernel with a $\sigma = 4$ pixels ($0.8 \, \rm arcsec$), without any additional smoothing along the wavelength axis in order to preserve the spectral resolution. During the extraction, we mask the positions of continuum sources to avoid contamination by positive or negative residuals from continuum subtraction. 
A three-dimensional group of connected voxels is classified as associated with the overdensity if its median velocity lies within $\pm 1000 \rm ~km\,s^{-1}$ of the mean redshift of the structure. This specific velocity window ensures we fully capture the velocity spread of the galaxies in the overdensity (up to $\approx 800-1000$ \kms, consistent with the velocity dispersions reported in Table \ref{tab:combined_overdensities}), while remaining narrow enough to exclude unrelated signal.
\begin{figure}
    \centering
        \includegraphics[width=\hsize]{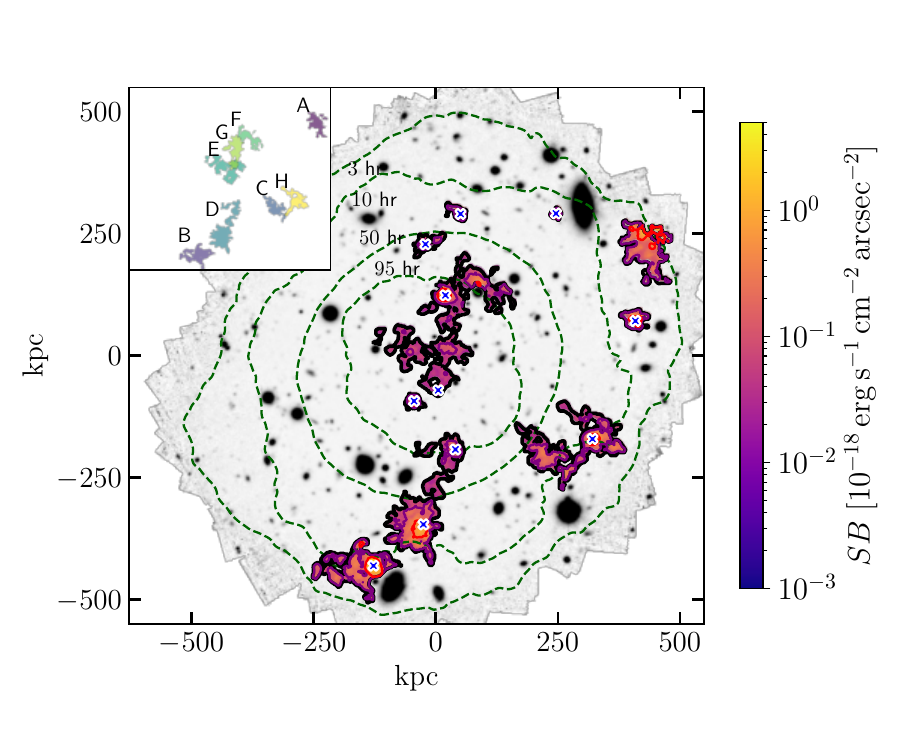}
        \vspace{-0.7 cm}
        \caption{Extracted \lya image of a cosmic web filament in the MUDF overdensity at $z \approx 3.047$. The black contour marks the $S/N = 2$ detection limit. The purple and red contours are at $1 \times 10^{-19}$ and $5\times 10^{-19}$ \sblcgs. LAEs are marked with blue crosses. The background shows the MUSE white-light image of the MUDF, with continuum-detected sources visible in black, and the dashed green contours indicate the exposure-time map of the field, labeled accordingly. In the \textit{upper-left inset}, the different two-dimensional projected structures of the \lya signal, obtained through the extraction procedure described in the text, are shown, ordered by velocity and labeled accordingly (see Fig. \ref{fig:ex3047b}). Only structures more extended than LAEs are labeled, as the extraction algorithm is optimized for diffuse emission and does not recover compact sources.}
        \vspace{-0.5 cm}
        \label{fig:ex3047a}
\end{figure}

We further perform a series of tests to assess the significance of the identified detections. For each structure, we derive the spectrum from the final datacube using the aperture defined by the two-dimensional projection of the three-dimensional extraction map. 
Comparing the extracted emission spectrum with the $1\sigma$ noise level computed with $3$-sigma clipping in the spectral region outside $\pm 1000$ \kms, allows us to evaluate the significance of each detection and to exclude cases that are consistent with background fluctuations. We also ensure that each detection does not overlap, within the velocity window, with any other low-redshift line emitters (e.g., [\ion{O}{II}], \ion{C}{III}], [\ion{O}{III}], \ion{H}{$\beta$}) identified by MUSE and classified in our catalog.

In addition, we test the extraction procedure on five random spectral slices, located far from the known overdensities, and on the inverted (i.e., multiplied by $-1$) version of the datacube. These tests provide a blind assessment of the algorithm's behavior and of the background properties at different redshifts. In these control runs, we do not recover any connected group of voxels that can be attributed to signal with an identifiable line-shape profile -- or any group of voxels at all -- reinforcing the robustness of the extraction procedure adopted for the overdensities. Finally, we verify that consistent results are obtained by applying a multiscale analysis using a wavelet-based approach, as discussed in \cite{Bacon2021} (see Appendix \ref{apx:wavelet}).

\begin{figure}
\includegraphics[scale=0.55]{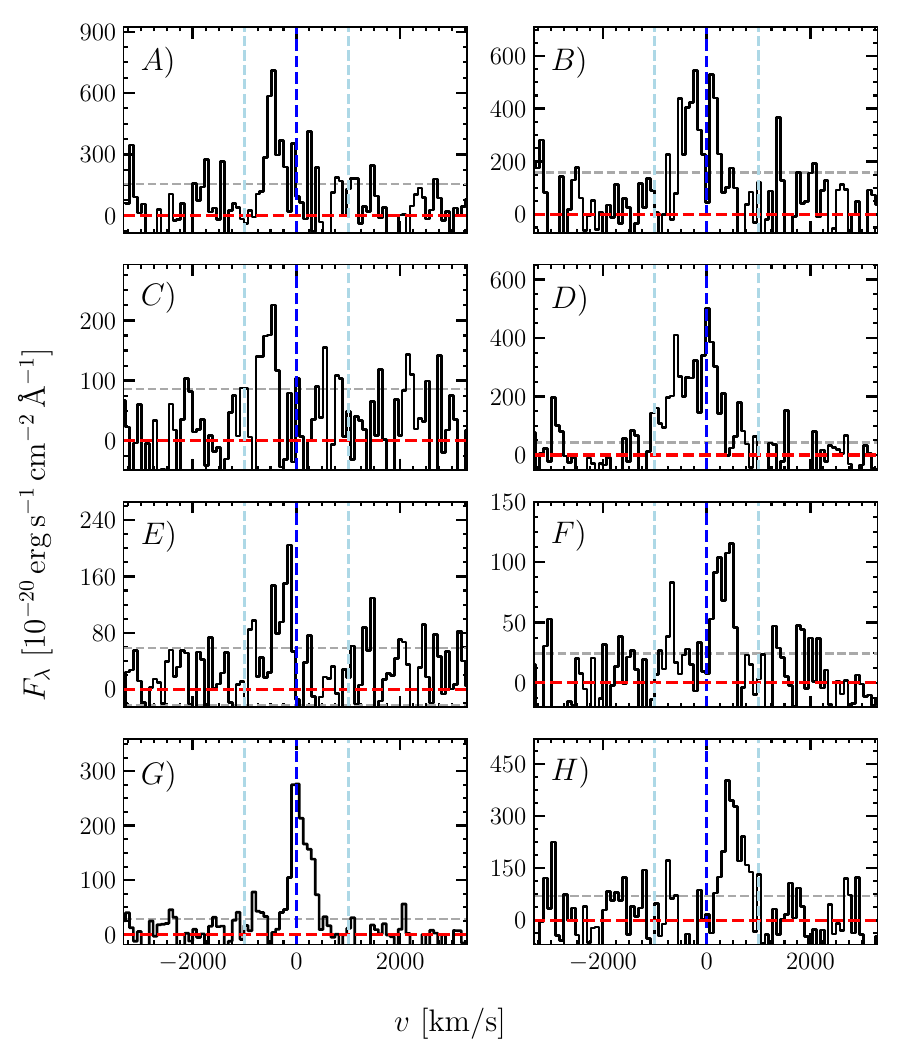}
        \caption{Extracted spectra from each emitting structure labeled in Fig.~\ref{fig:ex3047a} (black). The dashed red horizontal line marks the zero background level, while the dashed blue vertical line indicates the zero reference velocity computed from the mean redshift of the LAE overdensity. The vertical dashed light-blue lines show the $\pm 1000 \rm \, km\,s^{-1}$ velocity interval within which the identified structures are associated with the overdensity. The horizontal dashed gray line represents the $1\sigma$ noise level.}
        \label{fig:ex3047b}
\end{figure}

\begin{figure*}
\centering
\includegraphics[width=\hsize]{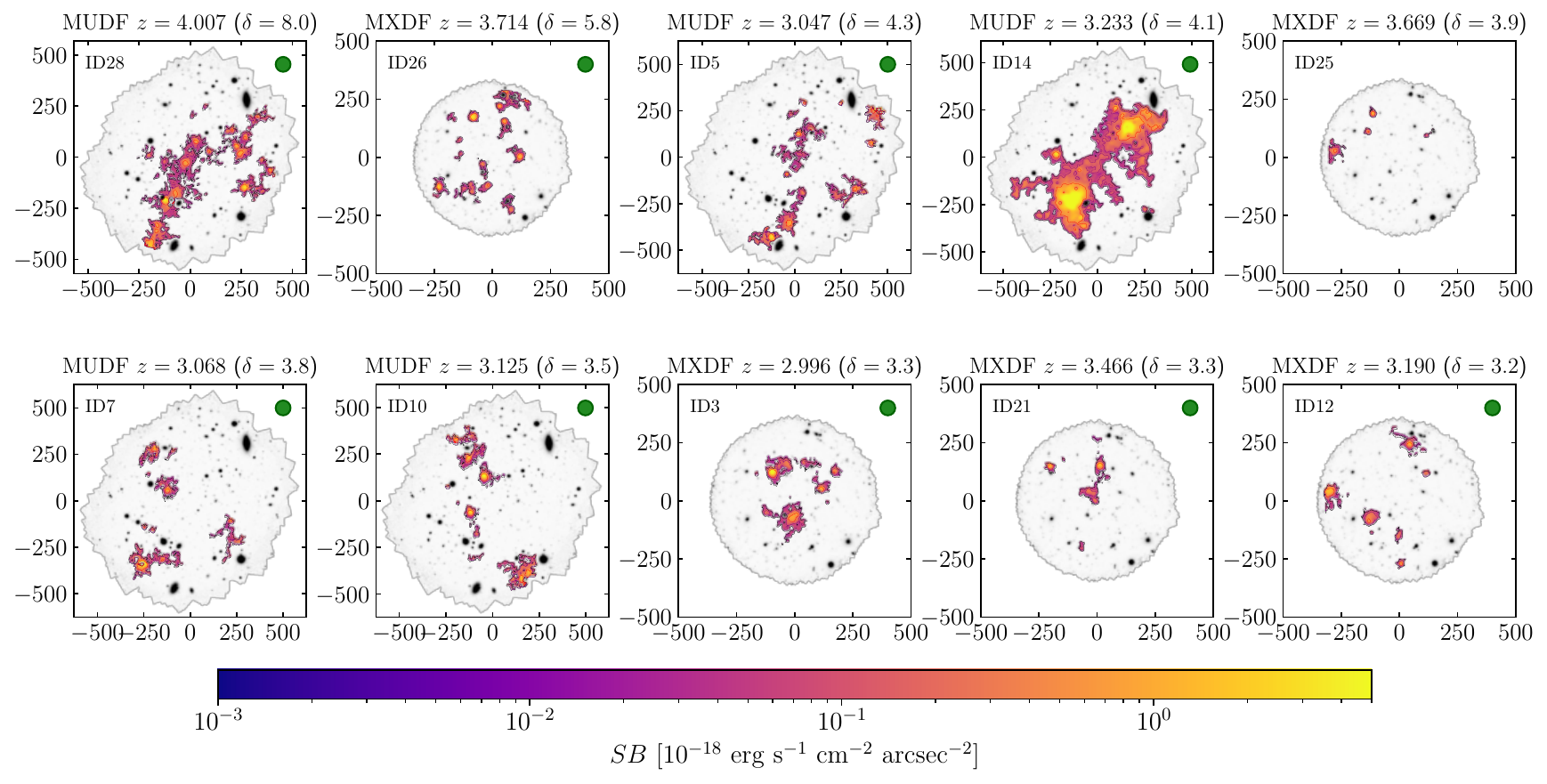}
\caption{\lya SB maps for the ten LAE overdensities with the highest overdensity value $\delta_\mathrm{LAE}$, sorted by decreasing $\delta_\mathrm{LAE}$ (see the text for the extraction details). The emission maps are overlaid on the white-light images of the corresponding survey footprints (MUDF or MXDF). Overdensities classified as hosting extended diffuse emission are marked with a green dot in the upper corner of each panel. The remaining overdensities are shown in Fig.~\ref{fig:gallery-full}.}
\label{fig:gallery}
\end{figure*}

The final validated signals are then collapsed along the wavelength axis to produce a two-dimensional \lya emission map \citep[as described in][]{Tornotti2025b}, highlighting the morphology of the detected emission. We note that the identified emitting structures generally contain at least one LAE embedded, with only a few cases showing no evidence of associated compact galaxy emission. In cases where no extended emission is associated with LAEs, the extraction map contains only the compact emission of these sources. An example of the output from this extraction and validation procedure is shown in Fig.~\ref{fig:ex3047a} and \ref{fig:ex3047b}, while all the resulting emission maps and associated spectra are presented as online material. In Fig. \ref{fig:gallery}, we report the emission maps of the ten overdensities with the highest $\delta_\mathrm{LAE}$, while the gallery of the remaining overdensities, sorted by decreasing overdensity, is shown in Fig. \ref{fig:gallery-full}. For the previously investigated overdensities at $z\approx 3.233$ and $z\approx 4.007$, the reader can refer to \cite{Tornotti2025c}\footnote{Here, we explicitly display the emission surrounding the LAE close to the main filament in the emission map.} and \cite{Tornotti2025b}, respectively.

We highlight a crucial observational limitation: the overdensity ID 24 and those at $z\gtrsim 4.4$ (ID 35, 36, 37, 39, and 40) reside in a wavelength region strongly affected by multiple nearby skylines (see Fig.~\ref{fig:z-distr}). This high sky-background contamination makes the search for extended, diffuse emission extremely challenging, limiting our detection capacity to only the bright, compact \lya line emission from the LAEs in these higher redshift environments. Consequently, the effective sample available for a robust analysis of extended \lya emission is reduced to $35$ systems.

\section{Analysis of the filament sample} \label{sect:analysis}
\subsection{Characterization of the diffuse emission}
Previous studies of cosmic web filaments detected in \lya emission -- both those connecting the bright quasar pair at $z\approx 3.22$ \citep{Tornotti2025c} and the large overdensity of LAEs at $z\approx4$ \citep{Tornotti2025b} in the MUDF -- have revealed a similar intrinsic SB in the intergalactic regions of these structures. This is particularly intriguing given the very different environments in which they reside: the former close to two powerful quasars, and the latter located in a galaxy overdensity with no clear signatures of AGN activity capable of significantly boosting the \lya emission. 
These results pose the question of how the observed properties of extended \lya structures -- such as their mean SB -- depend on the underlying galaxy overdensity. However, testing this connection depends on our ability to disentangle the contribution to the observed signal from the CGM of the embedded galaxies from that of the IGM, where no galaxies are detected at the depth of our data. 
A robust separation of CGM and IGM emission is therefore necessary to assess whether the observed \lya structures are primarily shaped by local galaxy environments. 

Previous studies on clustering properties of LAEs \citep[e.g.,][]{Bielby2016, Ouchi2020, Herrera2025, Umeda2025} placed them in dark matter halos with mass $\approx 10^{11} M_\odot$ at redshift $z\approx 3-4$ \citep[see also][]{Bacon2021}, with an expected virial radius $R_\mathrm{vir}\approx 35$ kpc. As also found in \cite{Tornotti2025b}, the SB profile of stacked LAEs embedded in the filament of the $z\approx 4$ overdensity shows evidence for a transition radius which is in agreement with the expected virial radius of halos at this mass scale. 
Motivated by this observational evidence, we define the diffuse emission by excluding the projected CGM contribution of each observed LAE within a circular aperture of radius $R=35$ kpc from the full SHINE two-dimensional projected map. We acknowledge that this geometric approximation may introduce some contamination from asymmetric CGM emission into the diffuse component, or, in some instances, may slightly overestimate the excluded area. Nevertheless, this approach provides a good starting point for the CGM boundary of the population as a whole. For the quasar pair at $z=3.233$, we mask the CGM according to the transition radii measured from the SB profile along the cosmic web filament, specifically $117$ kpc for the southeastern quasar and $108$ kpc for the northwestern quasar \citep[see Table 1 in][]{Tornotti2025c}.
We do not classify as hosting extended diffuse emission those overdensities that, following CGM removal, retain a total diffuse area smaller than $50 ~\rm arcsec^2$ ($\lesssim 3\times10^3 ~\rm kpc^2$ at $z=3$). This threshold ensures a sufficiently large area for statistically robust measurements. The excluded systems are used only to provide upper limits on the properties of the diffuse emission. Consequently, our final sample consists of $26$ overdensities with detected diffuse emission and $9$ without (see Figs. \ref{fig:gallery} and \ref{fig:gallery-full}).

 \begin{figure}
\includegraphics[width=9cm]{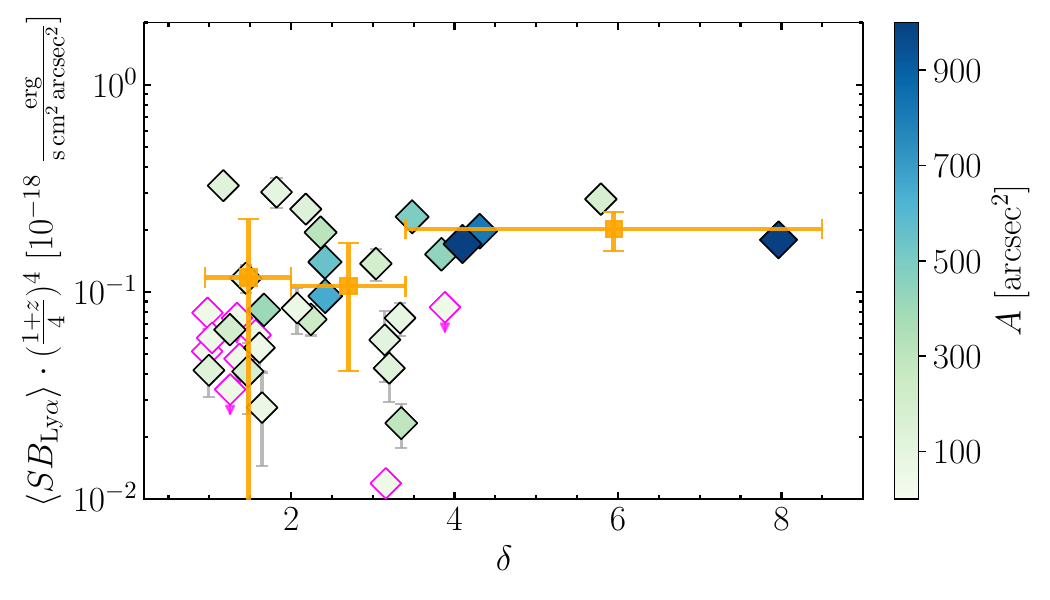}
\caption{Average intrinsic SB of the diffuse Ly$\alpha$ emission as a function of the local LAE overdensity. The vertical error bars display the variance-propagated $1\sigma$ measurement errors, and the magenta diamonds with arrows indicate the $1 \sigma$ detection limit of overdensities where no clear extended emission is found. The points are color-coded based on the area of the extended diffuse emission. The orange squares and associated error bars show the mean SB and standard deviation within the defined overdensity bins.}
\label{fig:sbint-ov}
\end{figure}

The three-dimensional extraction maps obtained with SHINE are optimized to detect low-SB emission and characterize the morphology of the extended structure. However, because these maps only sum flux from selected voxels along the wavelength dimension, they can systematically underestimate the total flux and are thus not suitable for estimates of the total SB \citep[e.g.,][]{Borisova2016, ArrigoniBattaia2019}. Therefore, to be consistent with what is usually done in the literature, we make use of a $30$~\AA-wide narrow-band image centered at the mean redshift of the extended structure for all subsequent measurements of SB and total flux. The SHINE projected map is then employed solely to define the spatial boundary (or morphology) for integrating the flux from the NB image while excluding, as described above, the contribution from the CGM. From these NB images, we measure the average intrinsic SB for the diffuse emission region, propagating the variance accordingly and correcting for cosmological dimming using the mean redshift of the overdensity. The resulting \lya intrinsic SB, $\langle \SB_\mathrm{Ly\alpha}^\mathrm{int}\rangle$, allows us to directly compare the observed properties of the extended gas across different cosmic times. 

By studying the dependence of $\langle \SB_\mathrm{Ly\alpha}^\mathrm{int} \rangle$ on the local LAE overdensity, we find in Fig.~\ref{fig:sbint-ov} a relatively flat trend, especially considering the highest overdensities ($\delta \gtrsim 3.5$). When binned into intervals of $\delta$ $[1-2]$, $[2-3.4]$, and $>3.4$ -- chosen to ensure a meaningful statistical number of detections per bin ($\geq 6)$, and to separate the regimes of low-, intermediate- and high-density -- the average SB is $\approx1 \times 10^{-19}$ \sblcgs\ for the lower-density regions and reaches $\approx 2 \times 10^{-19}$ \sblcgs\ at the highest overdensities.
To assess the robustness of our low-SB measurements, we quantify the SB limit for each NB image. We evaluate the distribution of average intrinsic SB values across $5000$ randomly positioned rectangular apertures, each having the same area as the identified diffuse emission region. The same procedure is performed for the overdensities classified as not hosting extended diffuse emission using the threshold area of $50 \rm ~arcsec^2$. These apertures are sampled only from the background area (defined by excluding the emitting regions and masking all known continuum sources). We define the detection limit as the $84^{\mathrm{th}}$ percentile of this background distribution ($1\sigma$ confidence level). Our measured average intrinsic SB values lie above the detection limits, typically at $\gtrsim 3\sigma$.

Finally, we explicitly address the possibility that \lya emission associated with the CGM may extend beyond the nominal virial radius. Stacking analyses of large LAE samples show radial profiles flattening at $\approx 20-50$ kpc and extending to $\gtrsim 100$ kpc \citep[e.g.,][]{LujanNiemeyer2022, Guo2024}. While this flattening can be interpreted as the transition from the CGM to the IGM, the interpretation of stacked profiles is complicated by the geometric dilution of signal coming from filaments below the detection limit and from genuinely empty regions, as discussed in \cite{Tornotti2025b}. Unlike stacking, our approach leverages individual detections to separate these components. By considering different CGM radii -- from $35$ kpc to $100$ kpc -- we find that the $\langle \SB_\mathrm{Ly\alpha}^\mathrm{int} \rangle$ of the remaining diffuse emission remains stable across the different overdensity bins. Specifically, we observe variations of $\approx 8-12 \%$ for $\delta \lesssim 3.5$, which drop to $\approx 4 \%$ for $\delta \gtrsim 3.5$. This stability holds despite the reduced area classified as diffuse and the increasing fraction of non-detections as the CGM radius increases. These results confirm that the observed trend in Fig. \ref{fig:sbint-ov} is robust against the specific CGM and IGM definition, and that the SB measured in the diffuse component is not driven by residual emission from the outskirts of individual galaxy halos. For this reason, we maintain the virial radius as the statistical definition for the CGM size, as justified above.

To investigate the structural connection between the galaxy distribution and the diffuse gas, we performed a Principal Component Analysis (PCA) on the spatial positions of the LAEs within each identified overdensity. This allows us to quantify the elongation and orientation of the overdensities through an alignment parameter $\xi_\delta$ (where $\xi_\delta=0$ corresponds to a random distribution and $\xi_\delta=1$ to a perfectly linear alignment). While we find no significant correlation between this parameter and the diffuse properties of the \lya emission (i.e., area or intrinsic SB), more than half of the overdensities are elongated, with a median alignment value of $\xi_\delta \approx 0.55$. 
This corroborates the hypothesis that LAEs lie along filamentary structures. 

\begin{figure}[t]
\centering
\resizebox{0.9\hsize}{!}{\includegraphics{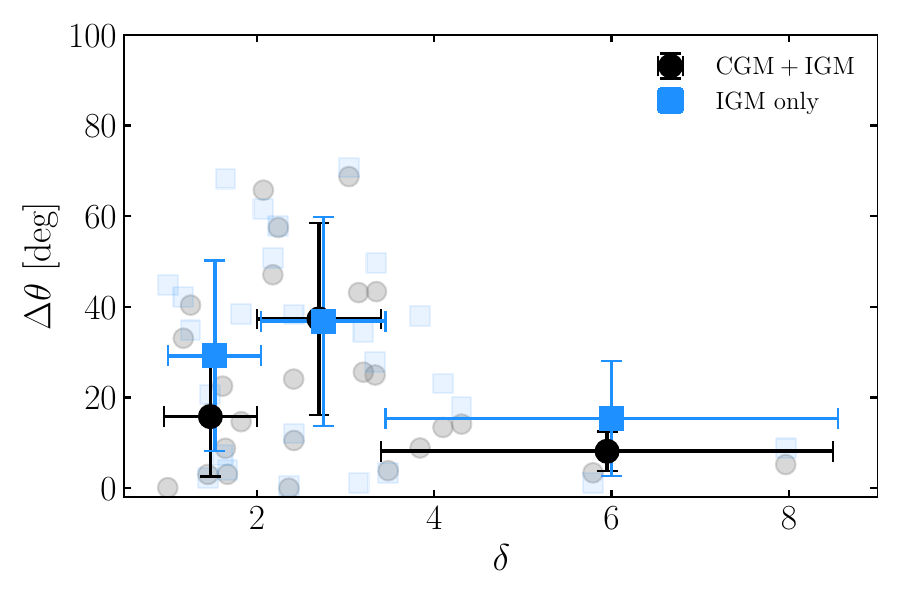}}
\caption{Misalignment angles between the major axis of the LAE distribution and the gas orientation, calculated for both the total emission (black) and extended diffuse component (blue), as a function of the overdensity. The average misalignment and standard deviation within the same overdensity bins defined in Fig. \ref{fig:sbint-ov} are shown. A slight offset of the bins is applied for better visualization.}
\label{fig:misalign-ov}
\vspace{-0.5 cm}
\end{figure}

We further compare the orientation angle of the galaxy distribution with the morphology of (i) the total emission (CGM plus extended filamentary diffuse gas) or (ii) the purely diffuse component. To determine this orientation angle, we treat the two-dimensional projected map of the emission as a spatial distribution. We first compute the spatial centroid (first-order image moments) to define the center of the structure. We then calculate the second-order central moments ($\mu_{20}, \mu_{02}, \mu_{11})$ of this spatial distribution, which effectively define its spatial covariance matrix. The orientation angle of the major axis is finally obtained from these central moments as $\theta = \frac{1}{2}\arctan[2\mu_{11}/(\mu_{02}-\mu_{20})]$, corresponding to the direction of the eigenvector associated with the largest eigenvalue. As shown in Fig. \ref{fig:misalign-ov}, the misalignment angles for both the total emission and the extended diffuse component follow a consistent trend. Although we observe a mean misalignment up to $\approx 35^\circ$ for $\delta \lesssim 3.5$, the majority of the overdensities are consistent with a scenario in which the gas follows the direction of the filaments as traced by the LAE positions within $\approx 20^\circ$. Crucially, even after the removal of the CGM contribution, the IGM remains closely aligned with the filamentary axis, particularly in the highest overdensities ($\delta > 3.5$), where the average misalignment angle is $\approx 15^\circ$. These results suggest that the diffuse gas generally traces the preferential axis defined by the member galaxies, reinforcing the idea that the selected overdensities mark the position of cosmic web filaments within which gas and galaxies are found.

\begin{figure*}
\sidecaption
\includegraphics[width=12cm]{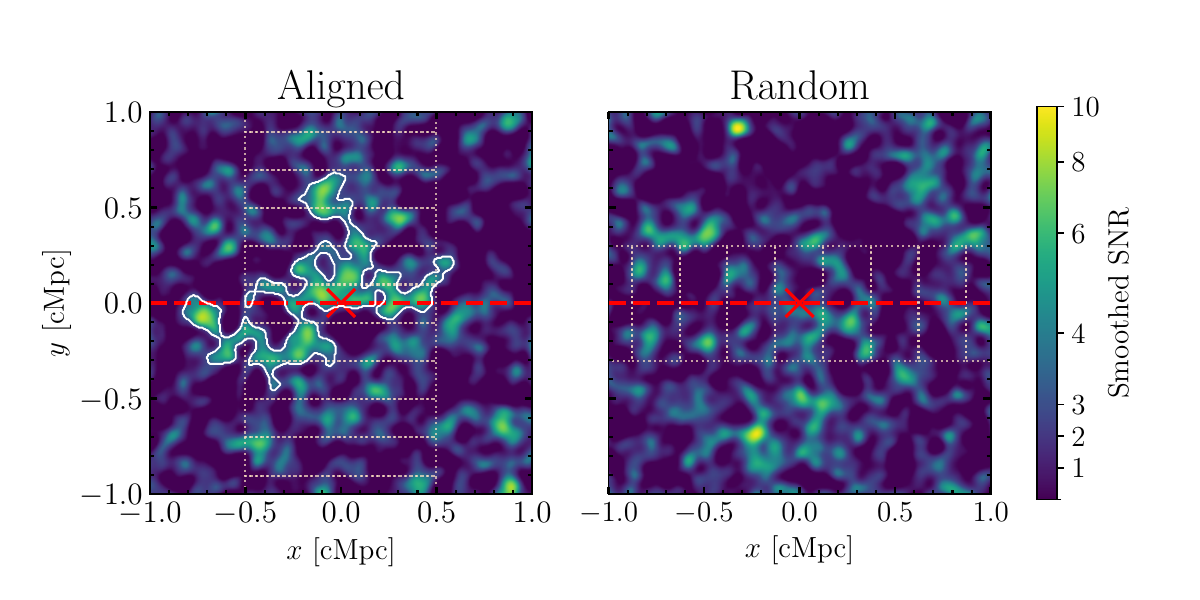}
\caption{Stack of the \lya emission in overdense environments ($\delta > 3.5$), once aligned along the direction of the cosmic web as traced by LAEs. The \textit{left panel} shows the smoothed ($1''$) $S/N$ map resulting from the oriented stacking procedure. The white contour indicates the $S/N=3$ boundary of the coherent structure extracted using the SHINE algorithm. The \textit{right panel} displays the same procedure applied with random orientations, serving as a null test to validate the significance of the aligned detection. In both panels, the dashed red line indicates the axis along which the images are aligned, and the cross denotes the common centroid. The dotted pink lines in the left and right panels show the boxes used to compute the perpendicular and horizontal SB profiles, respectively.}
\label{fig:stacking}
\end{figure*}

\begin{figure}[t]
\centering
\resizebox{0.95\hsize}{!}{\includegraphics{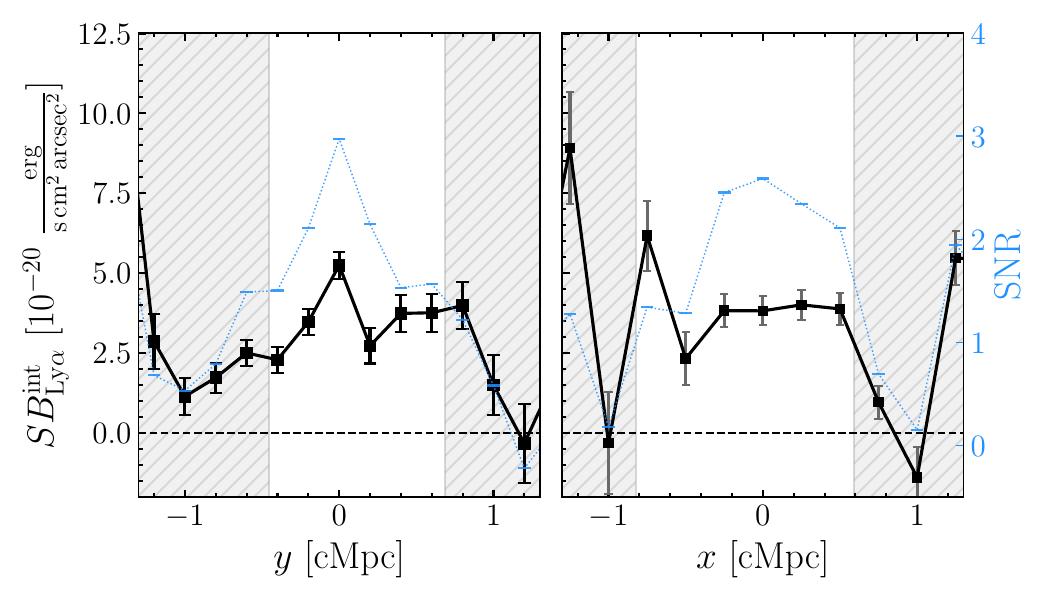}}
\caption{\lya SB profiles of the stacked high-overdensity sample. The \textit{left panel} shows the intrinsic SB profile (black) computed perpendicular to the filament axis, while the \textit{right panel} is along the alignment axis (see boxes in Fig. \ref{fig:stacking}). In both panels, the blue lines represent the corresponding mean smoothed $S/N$ profiles with reference to the panel's right-hand side axis. The gray shaded regions indicate the areas beyond the spatial boundaries of the coherent structure extracted with SHINE.}
\label{fig:sbprofiles-stack}
\end{figure}

\subsection{Stacking}

The tight alignment between diffuse emitting gas and the location of LAEs motivates an oriented stacking analysis of the subsample composed of high overdensities to further unveil the average emission properties of the cosmic web.
Within high-overdensity regions $\delta\gtrsim 3.5$, we observe that diffuse \lya emission consistently emerges with a maximum characteristic $\langle \SB_\mathrm{Ly\alpha}^\mathrm{int} \rangle \approx 2\times10^{-19}$ \sblcgs. To statistically investigate the distribution of this gas phase beyond individual detections, we implement an oriented stacking strategy applicable to both current and future wide-area surveys, where diffuse emission at these SB levels may remain below individual detection limits. 

We define the stacking axis based on the orientation angle derived from the Ly$\alpha$ CGM emission associated with the member LAEs (see previous paragraph). By aligning the major axis of the galaxy-CGM distribution, we establish a geometric framework for searching for the underlying intergalactic gas. This methodology remains robust even with shallower data where only the CGM is visible. To ensure the orientation accurately reflects the large-scale structure, we refine the axis definition in systems with a main alignment that may be skewed by a few peripheral members. To perform the stack, we employ the $30$~\AA-wide NB images for the subsample of high-overdensity systems ($\delta\gtrsim3.5$, 7 systems), scaled by the cosmological dimming factor and renormalized to $z=3$ using the factor $[(1+z_\delta)/4]^4$. To ensure the final stack reflects only diffuse emission, we mask all known continuum sources and the CGM contributions.
Each image is then rotated according to its determined orientation angle, centered on its centroid, and interpolated onto a common grid in comoving coordinates (ckpc). The specific orientation angles and centroids adopted for each overdensity are summarized in Table \ref{tab:stack_info}. We adopt a spatial sampling given by the MUSE pixel size in ckpc at $z=3$ ($\approx 6.3$ ckpc); the pixel value is assigned to the nearest new pixel to conserve the SB with respect to the initial image \cite[see, e.g.,][]{Gallego2018}. 
The final stack is produced by computing the average SB for each pixel, with the variance propagated accordingly.

This procedure results in a stacked image with maximum depth of $\approx 800$ hr in the central region, which decreases sharply toward the edges. To highlight significant emission arising solely from the diffuse component, we apply the SHINE algorithm to the stacked image using a Gaussian smoothing kernel of $5$ pixels, an $S/N$ threshold of $3$, and a minimum number of connected pixels of $8000$ ($\approx 3.7\times10^5 \rm~ckpc^2$). This configuration -- with a broader smoothing area but more stringent area and $S/N$ threshold -- is optimized for the very low-SB, large scale nature of the expected signal.
In the left panel of Fig.~\ref{fig:stacking}, we present the resulting stacked emission and the corresponding extraction overlaid on the smoothed $S/N$ map, derived from the smoothed stacked image and variance. We check that a similar result is consistently obtained using a median stack. However, we adopt the average as our reference to remain consistent with the physical interpretation presented in the following discussion (see Sect. \ref{sect:cov-fact}). Moreover, to assess the robustness of the alignment, we perform a control stack using randomized orientations. As shown in the right panel of Fig.~\ref{fig:stacking}, this produces no detectable coherent signal when performing the same extraction process. Additionally, we also verify that the same random stack produces a null detection when masking all known emission.

To characterize the emission properties of the identified structure within the stack, we measured the intrinsic SB profiles from the unsmoothed stacked image. We computed the mean SB using rectangular apertures aligned both parallel ($250\times600$ ckpc) and perpendicular ($1000\times200$ ckpc) to the expected filament axis, propagating the variance accordingly. To account for the varying depth across the field, we also derived corresponding profiles from the smoothed $S/N$ map. This analysis confirms a marginal but coherent excess of signal in the stack, with a mean intrinsic SB of $\langle \SB_\mathrm{int}^\mathrm{stack} \rangle\approx 4\times10^{-20}$ \sblcgs\ (see Fig. \ref{fig:sbprofiles-stack}).

\section{Discussion} \label{sect:discussion}

A coherent analysis of the two deepest MUSE datasets acquired to date, the MXDF and the MUDF, is enabling a first statistical investigation of the emission properties of the $z\gtrsim 3$ cosmic web.
With these data, we aim to investigate the following three questions: (i) whether LAE overdensities act as signposts of the cosmic web; (ii) whether the emission properties of the cosmic web change as a function of galaxy environment; and (iii) what we can infer about the underlying physical conditions of the cosmic web based on a comparative statistical analysis of the \lya emission in filaments. 

\subsection{Statistical properties of the cosmic-web \lya emission}

By homogeneously leveraging a sample of $41$ LAE overdensities in the redshift range $z\approx 3-5$, we demonstrate that extended diffuse \lya emission is detectable beyond the CGM, especially in high-density environments ($\delta \gtrsim 3.5$). For $\approx 74\%$ of the sample, we detect extended emission outside the CGM, in areas $\gtrsim 50 \rm \, arcsec^2$. This fraction rises to $86\%$ when considering overdensities with $\delta \gtrsim 3.5$.
In several cases ($8/35$), this emission hints at the presence of large-scale structures or directly resembles cosmic web filaments.
Moreover, a statistical analysis of the preferred alignment of diffuse gas reveals how \lya emission is, on average, aligned in the direction of a preferential axis as traced by LAEs. This is especially true for larger overdensities, where no cases of isotropic emission are identified. Combining this empirical evidence, we derive the first result of our statistical analysis: at $z\gtrsim 3$, rich LAE overdensities (particularly with $\delta \gtrsim 3.5$) serve as signposts of large-scale ($\gtrsim 1-2$ cMpc) filaments \citep[see also][]{Bacon2021}. 

As shown in Fig. \ref{fig:sbint-ov}, the diffuse gas detected above our sensitivity limit maintains a remarkably consistent SB level of $\SB_\mathrm{Ly\alpha}^\mathrm{int} \approx 2\times10^{-19}$ \sblcgs\ with no strong dependence on the overdensity, especially for $\delta>3.5$ \citep{martin_extensive_2023}. 
Instead, it seems that the scatter in this relation and the fraction of non-detections are more dependent on overdensity. 
Most non-detections are found at $\delta <2$, in the interval $2\lesssim \delta \lesssim 3.5$, we observe a $\approx 1$~dex downward scatter compared to the maximum SB level. Despite the limited statistics, at $\delta \gtrsim 3.5$ the values instead reach this asymptotic level.  As already noted in \citet{Tornotti2025b}, this maximum SB also appears constant regardless of the presence of bright AGNs. The most striking example of this behavior is the filament connecting the two bright, $z\approx 3.22$ quasars in the MUDF \citep{Tornotti2025c}, where the emitting diffuse gas in the filament has SB levels in line with the rest of the sample (see the fourth panel in the top row of Fig. \ref{fig:gallery}).

This nearly constant SB suggests a possible asymptotic behavior for the \lya emission of diffuse gas inside the cosmic web. Further support for this trend comes from the LAE overdensity at $z\approx 3.577$ reported by \cite{banerjee_2024}. Although observed with a shorter MUSE integration time (10 hours) that reveals patches of emission clustered to galaxies, we estimate that this filament lies inside an overdensity of $\delta \approx 12.6$, using the procedure described in Sect.~\ref{sect:over-def} and adopting a similar UDF-MOSAIC selection function. Despite the lower sensitivity in this field compared to other ultra-deep data, this system also exhibits \lya emission that can be ascribed to a filament of linear size $\approx 260$~kpc, with SB levels of $\approx 10^{-19}$~\sblcgs.

These pieces of evidence allow us to address the second science question above. \lya emission from cosmic filaments appears to vary with the overdensity parameter: low-to-mild overdensities host a wide range of filament brightnesses, which are typically fainter than the maximum value. Higher overdensities, instead, host brighter filaments, up to the maximum SB of $\SB_\mathrm{Ly\alpha}^\mathrm{int} \approx 2\times10^{-19}$ \sblcgs.
Addressing the third question about the physical conditions of the emitting gas is less straightforward, as this inference problem requires not only a detailed radiative transfer calculation \citep[e.g.,][]{Byrohl2023}, but it is also dependent on several physical parameters that cannot be uniquely constrained by the single observable we have, i.e., the \lya SB.  In the spirit of making some progress in addressing this question, however, we can follow simple lines of argument, in light of what has been proposed by \citet{Bacon2021}, \citet{Tornotti2025c}, and \citet{Tornotti2025b}.

In a simple scenario, \lya photons are produced by two emission mechanisms: recombination radiation following photo- and collisional ionization, and collisional excitation by free electrons \citep[e.g.,][]{Gould1996,faucher-giguere_ly_2010,Bacon2021,Byrohl2023}. 
The emissivity for both processes scales with the square of the density, and we can therefore explain the observed scatter and the increasing SB with overdensities through variation in the physical density. In fact, a mild variation of the gas density with $\delta$, within a factor of $\approx 3$, is sufficient to account for $\approx 1$~dex variation in the observed SB. 
Moreover, the emitting gas is most likely to be at temperatures around $T\approx 10^4$~K. This is the typical temperature of the \lya forest, as observed \citep{pettini_lyman-alpha_1990,Rudie2019} and predicted by numerical simulations \citep{theuns_p3m-sph_1998}.  Furthermore, optically thick gas clouds that give rise to Lyman limit systems (LLSs) with $\log(N_\mathrm{HI}/\rm cm^{-2})>17.2$ are also found in a comparable temperature range \citep{fumagalli_physical_2016}.
Around $T\approx 3\times 10^{4}$~K, the contribution of collisional excitation to \lya production becomes significant relative to radiative recombination for densities $\gtrsim 10^{-3}~\rm cm^{-3}$ \citep{faucher-giguere_ly_2010}, and it is therefore plausible that our observations, which are inevitably probing the denser and brighter patches of emission, preferentially select the inner spines of the filaments where collisions are important \citep{Byrohl2023}.
Following this line of thought, we can therefore constrain a lower limit on the density of the medium. The gas needs to be ionized but dense enough to account for the appreciable SB we observe, also via collisional excitation ($\log({N_\mathrm{HI}/\rm cm^{-2})} \gg 14-15$). 

A plateau in the observed SB with galaxy overdensity could initially be explained by a transition from the optically thin to the optically thick regime. However, if the gas is turning predominantly neutral and the emission is dominated by a skin of ionized gas that recombines, we would recover the well-known effect of a SB that depends on the impinging radiation field \citep{hennawi_quasars_2013}. This scenario seems, however, disfavored by the fact that the observed emission appears not to depend on the presence or lack thereof of very bright sources such as quasars, as noted in \citet{Tornotti2025b}. This places an upper limit on the gas density: the density cannot be too high for the gas to completely self-shield ($\log({N_\mathrm{HI}/\rm cm^{-2})} \ll 19-20$). 

Combining these arguments, we are left with the hypothesis that we are observing gas that is in the transition regime between ionized and neutral, a regime that is occupied by partially ionized gas that gives rise, when observed in absorption, to (partial) LLSs \citep{fumagalli_absorption-line_2011,faucher-giguere_small_2011,rahmati_distribution_2015}. 
Unfortunately, this regime is among the most difficult to capture with simple analytic descriptions or considerations and requires detailed radiative transfer calculations. Moreover, reaching SB as high as $\approx 1-2\times 10^{-19}$~\sblcgs\ is challenging in this density regime by including only recombination and collisions \citep{Byrohl2023,Bacon2021}. 
The processes described here must provide only a lower limit on the \lya photon budget, and scattering from galaxies \citep{Byrohl2023} or even the CGM of individually unresolved galaxies \citep{Bacon2021} becomes a needed contribution.  
The third question, about what constraints we can place on the physical properties of the emitting gas, remains unanswered, but this analysis demonstrates the potential of a comparative analysis of a first statistical sample of filaments to address this open issue.

\subsection{The covering factor of \lya emitting gas} \label{sect:cov-fact}

With a first sample of cosmic web filaments, we can build insight into the covering factor of emitting gas. This quantity is of interest for two reasons. First, we can relate the covering factor of the emitting gas with the cross section of the gas probed in absorption, to further constrain the underlying physical properties of the cosmic web. This discussion is presented in the following Sect. \ref{sect:incidence}. 
Second, and the focus of the current section, we can estimate the extent to which oriented stacking of putative filaments traced by galaxies provides an efficient way to unveil the average SB of the filaments. This is a worthwhile exercise, particularly because it will be difficult to grow the current samples of ultra-deep observations until new-generation instruments come online (e.g., the wide-field spectroscopic telescope, WST; \citealt{bacon_wst_2024}). In the medium-term, stacking will be our best way to increase sample size. 

\begin{figure}[t]
\centering
\resizebox{0.7\hsize}{!}{\includegraphics{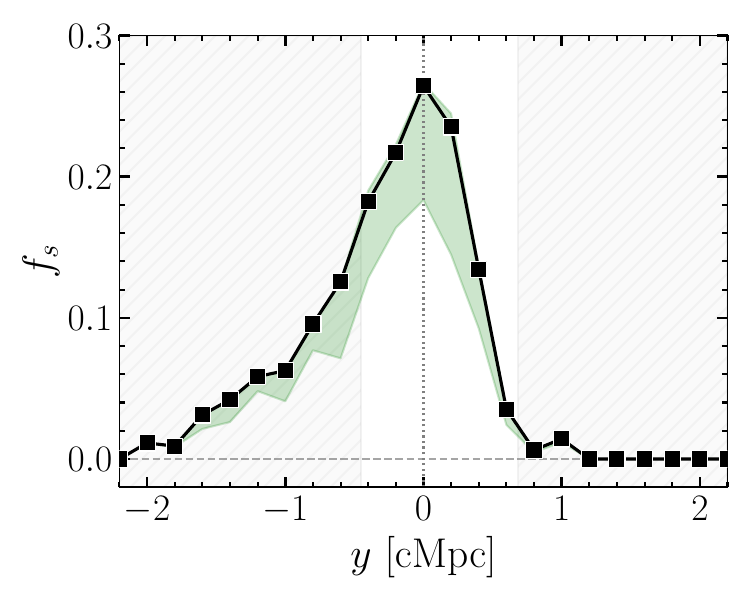}}
\caption{Covering factor profile of the aligned stack. The profile is computed within the same rectangular apertures used for the vertical SB profile (see left panel of Fig. \ref{fig:stacking}). For each pixel, the covering factor is defined as the fraction of systems contributing detected emission to the total number of stacked overdensities ($N=7$). The black line and the green shaded region represent, respectively, the mean and the 16th–84th percentile range derived from a jackknife resampling experiment. Gray shaded regions indicate the areas outside the spatial boundaries of the coherent structure identified by the SHINE algorithm.}
\label{fig:covfac}
\end{figure}

Despite the limited statistics, the seven overdensities at $\delta > 3.5$ provide compelling motivation for testing how a stacking analysis (Fig.~\ref{fig:stacking}) can constrain the spatial distribution of the covering factor of the emitting gas. Assuming that the gaseous filaments are correctly aligned along their major axes by our procedure, any decrease in $\SB_\mathrm{int}^\mathrm{stack}$ from the stack of $N_\mathrm{tot}$ filaments, compared to the maximum $\SB_\mathrm{Ly\alpha}^\mathrm{int}$, can be interpreted as signal dilution due to the spatial covering factor, $f_s$:
\begin{equation}\label{eq:stacksb}
    \SB_\mathrm{int}^\mathrm{stack}= \frac{1}{N_\mathrm{tot}}\sum_{i=1}^{N_\mathrm{tot}}\SB_{\mathrm{int},i} \approx f_s\times \SB_\mathrm{Ly\alpha}^\mathrm{int}\:.
\end{equation} 
We define $f_s$ as the fraction of overdensities contributing to the detected emission in each pixel in the stack or, equivalently, as the probability of intersecting emitting gas along a line of sight that goes through a filament. 
This quantity can be measured directly in our sample by counting the fraction of pixels covered by emission at the sensitivity of our data. As shown in Fig.~\ref{fig:covfac}, the average $f_s$ profile measured within the rectangular apertures used to measure the vertical SB profile in the stack (see left panel of Fig. \ref{fig:stacking}) reaches values of $\approx 0.20 - 0.25$ at the expected filament positions \citep{gallego_constraining_2021}. We rely on the vertical profile (perpendicular to the alignment axis) as it physically probes the transverse cross-section of the filaments, characterizing the decline of the gas covering factor from the central spine outward.

From this direct measurement of $f_s$, we can relate the stacked SB ($\SB_\mathrm{int}^\mathrm{stack}$) to the individual contributions of each filament ($\SB_{\mathrm{int},i}$) across $N_\mathrm{tot}$ images through Eq.~\ref{eq:stacksb}.
In the $(1 - f_s)$ fraction of cases where no emission is significantly detected, the stack incorporates regions that are noise-dominated. While these regions likely contain faint Ly$\alpha$ emission from the underlying cosmic web, this signal remains below our current sensitivity. Stacking reduces the stochastic noise, but we can assume that the contribution of this faint and more extended component remains negligible compared to the signal originating from the $f_s$ fraction of the pixels that contain brighter emission. 
The net effect is a dilution of the observed SB in the stack compared to the maximum value. 
This is indeed what we observe: the geometrically measured $f_s\approx 0.20-0.25$ combined with the observed $\langle \SB_\mathrm{Ly\alpha}^\mathrm{int} \rangle \approx 2\times 10^{-19}~$\sblcgs\ yields an expected $\langle \SB_\mathrm{int}^\mathrm{stack} \rangle \approx 4-5 \times 10^{-20}$ \sblcgs, which is in good agreement with our measured stacked SB (see Fig.~\ref{fig:sbprofiles-stack}). While a direct one-to-one comparison is difficult, these SB values are in line with the levels detected at distances $\gtrsim 100$ kpc in the stacking analysis of 15 quasars at $z\sim2$ \citep{ArrigoniBattaia2016} and around bright LAEs ($\log(L_\mathrm{Ly\alpha}/\rm erg\,s^{-1})\gtrsim 42.4$) in \cite{LujanNiemeyer2022}. When rescaled to $z=3$ to account for cosmological dimming, these literature values are a factor of $\approx 2$ lower than our measurement. This discrepancy can be attributed to the combined effect of geometric dilution inherent in circular stacking across large radial bins and an intrinsically lower emissivity at lower redshifts, likely driven by the decrease in average gas density of the IGM.

\subsection{Comparing the incidence of \lya emission and optically thick gas} \label{sect:incidence}

In the previous sections, we have leveraged the statistical information from the analysis of our sample of emitting cosmic-web filaments to conclude that the emitting gas is likely in a range of densities typical of the transition between optically thin and optically thick gas, a physical state that is typically of LLSs. We also inferred that the emitting regions have a covering factor along a line of sight of $\approx 0.20-0.25$. In this section, we can combine these two findings to further learn about the ionization state of the gas by comparing, on purely statistical grounds, the probability of intersecting emitting gas and LLSs. A comparable incidence would further reinforce the inference above, while highly discrepant results would impose new constraints on the allowed range of gas densities. 

The first step in this comparison is to measure the cosmological incidence of \lya emission from the cosmic web. 
In analogy with what is performed by absorption line measurements, we consider the volumes probed by the MUDF and MXDF cubes between $z\approx 2.8-4.0$ and derive the incidence $\ell_{\rm Ly\alpha}$ by summing the emitting structures intersected along the probed redshift path normalized by the total path probed, $\ell(Ly\alpha) = N_{\rm structures}/\Delta z$.
In practice, we construct a masked cube where voxels containing detected emissions are assigned a value of unity, and non-detections are left at zero value. Sight lines toward bright continuum sources, or missing data along the spectral direction from VLT lasers or bright skylines are removed from the calculation. We then sum all values along the redshift axis, normalizing by the total redshift path probed. 

As this calculation depends on the sensitivity of each voxel, formally, we would need to account for a surface-brightness dependent completeness factor, as done, for instance, in the formalism to compute the incidence of 
absorption line systems \citep{hasan_evolution_2020}. 
However, we find that the measured incidence converges to within 10 percent once we restrict to voxels with an exposure time $\gtrsim 50$~hours. 
The recovered incidence is $\ell(\rm Ly\alpha) \approx 1.40$ in the MUDF and $\ell(\rm Ly\alpha) \approx 1.05$ in the MXDF. The two values underscore a non-negligible cosmic variance; hence, the quoted values should be interpreted with some large systematic uncertainty. Nonetheless, the combined MUDF+MXDF value of $\ell(\rm Ly\alpha) \approx 1.22$ is in the realm of the incidence of optically thick gas giving rise to LLSs measured in large quasar surveys, $\ell(\rm LLS) \approx 1.2-2.0$ between $z\approx 3-4$, with a typical value of $\ell(\rm LLS) \approx 1.5$ at $z\approx 3.5$ \citep{fumagalli_dissecting_2013,prochaska_definitive_2010,fumagalli_detecting_2020}. The incidence we derive for emitting gas with SB $\gtrsim 5\times 10^{-20}$~\sblcgs\ is also consistent with the cross-section of extended emission around LAEs as measured in deep stacks by \citet[][see their figure 4]{wisotzki_nearly_2018}. By following similar statistical arguments, they also relate this gas to LLSs with column density $\approx 10^{18}$~cm$^{-2}$ \citep[see also][]{gallego_constraining_2021}.   
Finally, \citet{martin_extensive_2023} argues for a correlation in the emission from filaments and the gas seen in absorption that extends to the lower column density \lya forest. While the argument they present shares many similarities with our analysis, they infer a much higher cross-section of bright emission than what we observe (see their Figure 5). However, they target rich overdensities known to harbor $z\approx 2.3-2.5$ protoclusters, which are non typical compared to the untargeted search we conduct in this study. Indeed, previous MUSE observations in protoclusters \citet{Umehata2019} indicate a $\approx 1~$dex brighter emission in the core of these extreme overdensities, an effect that can be attributed to elevated local radiation fields in an optically thick medium. Hence, we do not regard this discrepancy as a significant tension of our conclusions.    

We believe the emerging consistency between the emission from filaments and LLSs is not purely coincidental. Had we found a much larger incidence for the emitting gas, we would have been forced to conclude that a substantial fraction of the emitting gas is at a much lower density, typical of the \lya forest in the optically thin regime. Conversely, had we found a much smaller incidence of the emitting gas compared to LLSs, we would have needed to invoke gas with a smaller cross section, such as the neutral gas giving rise to high-column density damped \lya systems. The finding of a comparable incidence for the emitting gas and for LLSs reinforces, instead, the general considerations reported at the beginning of this discussion: we are observing in emission gas that is partially ionized at moderate densities, $n_{\rm H}\approx 10^{-3}-10^{-1}~\rm cm^{-3}$ \citep{Tornotti2025c,fumagalli_physical_2016}, with the potential contribution of embedded substructures. This gas likely originates from a combination of the CGM of galaxies and the denser spines of the filaments, which are contributing to both the observed LLSs in absorption and the brightest patches of emission we detect at the current depth of our MUSE observations \citep{Bacon2021}. This emerging picture is also consistent with the predictions of numerical simulations \citep{Byrohl2023,fumagalli_absorption-line_2011,faucher-giguere_small_2011}. 

\subsection{Forecast for future detections of intergalactic gas}
Expanding these ultra-deep samples through new dedicated observations is a challenging task. To make progress, it is therefore essential to leverage the already existing medium-depth MUSE datasets.   
Building upon the oriented stacking methodology described in the previous sections, we provide a simple forecast for the feasibility of future detections. Specifically, we predict the minimum number of overdensities, $N_{\mathrm{stack}}$, required in a stack to detect \lya from the intergalactic gas.
This number depends on two variables and three parameters, according to the following equation:
\begin{equation}\label{eq:Nstack}
    N_\mathrm{stack}(\sigma^\mathrm{int}_\mathrm{rms}) = \left( \frac{\sigma^\mathrm{int}_\mathrm{rms} \cdot n_{\sigma}}{f_s \cdot \SB_\mathrm{int}^\mathrm{max} \cdot \sqrt{A_\mathrm{K}} }\right)^2\:.
\end{equation}
The first parameter, $\SB_\mathrm{int}^\mathrm{max}$, is the expected SB in a system, which we assume from our observations in $\delta \gtrsim 3.5$ overdensities to be $\SB_\mathrm{int}^\mathrm{max} \approx 2\times10^{-19}$~\sblcgs. 
The second and third parameters relate to the detection experiment and are, respectively, the significance $n_\sigma$ with which the signal is detected in a reference aperture of $1~\rm arcsec^{2}$ \citep[e.g.,][]{ArrigoniBattaia2019}, and the effective area of the smoothing kernel normalized to the reference aperture, $A_{\mathrm{K}}$. Here, we take $n_{\sigma}=2$ and $A_{\mathrm{K}}=13$. This choice of $A_{\mathrm{K}}$ mimics an averaging filter of width $w$ equivalent to the typical Gaussian smoothing with $\sigma = 1''$ often employed in NB detection, following the relation $A_{\mathrm{K}} = w^2 = 12\sigma^2 + 1$ (with $\sigma$ in arcsec; e.g., \citealt{Wells1986}).
The two variables are the sensitivity of the image, $\sigma_\mathrm{rms}$, and the covering factor $f_s$. 
Basically, Eq.~\ref{eq:Nstack} is obtained imposing that the diluted signal $\SB_\mathrm{int}^\mathrm{stack} \approx f_s \cdot \SB_\mathrm{int}^\mathrm{max}$ in a stack is detected with a significance $n_\sigma$ once the noise is rescaled as $\sigma_{\mathrm{rms}}^{\mathrm{int}} / \sqrt{N_{\mathrm{stack}} \cdot A_{\mathrm{K}}}$, assuming it is Gaussian. We note, however, that potential non-Gaussian systematics would affect the noise scaling; therefore, this forecast represents a lower limit on the required $N_\mathrm{stack}$. 
\begin{figure}[t]
\centering
\resizebox{0.9\hsize}{!}{\includegraphics{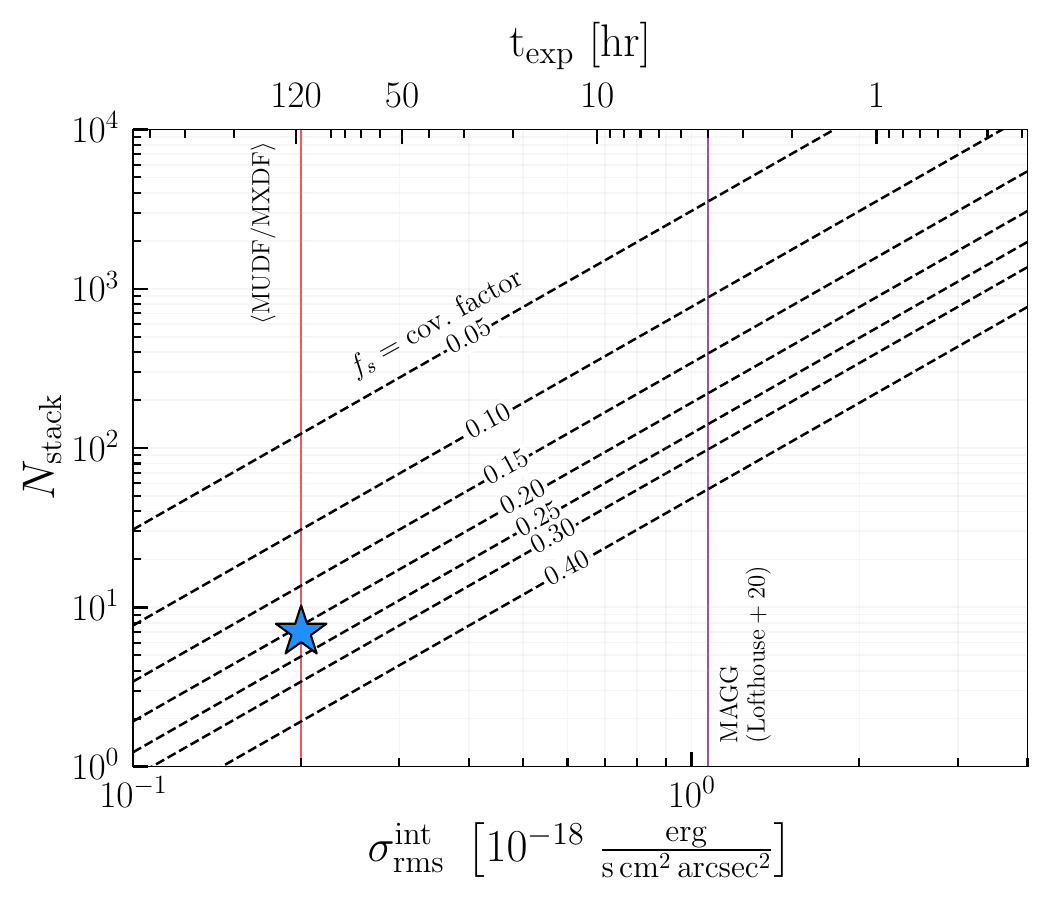}}
\caption{Predicted number of high-overdensity systems ($\delta \gtrsim 3.5$) required to detect the diffuse gas component in an aligned stack as a function of the individual image noise level. The calculation assumes a target emission at the average level $\langle \SB_\mathrm{Ly\alpha}^\mathrm{int} \rangle \approx 2 \times 10^{-19}$ \sblcgs\ for different covering factors, $f_s$. The x-axis represents the $1\sigma$ SB sensitivity ($\sigma_{\mathrm{rms}}$) for a $30\,\text{\AA}$-wide NB image -- corrected for cosmological dimming and rescaled to $z=3$ -- within a $1\,\mathrm{arcsec}^2$ aperture. Vertical solid lines indicate the average $\sigma_{\mathrm{rms}}$ for the MUDF+MXDF (red) and MAGG \citep[purple,][]{Lofthouse2020} datasets. The blue star denotes the result obtained from our MUDF+MXDF stacked subsample ($N=7$), which is consistent with a covering factor $f_s \approx 0.20-0.25$.}
\label{fig:prediction}
\end{figure}
In Fig. \ref{fig:prediction}, we show the expected 
number of overdensities to be included in a stack as a function of the depth of individual exposures (corrected for cosmological dimming and rescaled to $z=3$) and for a range of covering factors $0.05-0.40$.
A larger number of overdensities is required when using shallower data, and when the covering factor is lower, due to the geometric dilution in Eq.~\ref{eq:stacksb}.
As a consistency check of this forecast, we show that a stack of seven systems in our high-overdensity sample characterized by an average $\sigma_{\mathrm{rms}}^{\mathrm{int}} \approx 2 \times 10^{-19}$ \sblcgs\ leads to a detection when covering factors are $f_s \approx 0.20-0.25$. This is in line with the independent inference above. In Fig.~\ref{fig:prediction}, we also show the typical depth of shallower but larger surveys, such as the MAGG dataset \citep{Lofthouse2020}. For $f_s \approx 0.25$, a stack of $\approx 150$ overdensities observed at medium depths is expected to yield a detection. Assembling a sample of this many overdensities is within reach, given the current status of the MUSE archive. 

However, besides the assumptions made for the parameters of Eq.~\ref{eq:Nstack}, a caveat also applies to the identification of the alignment axis. Shallower data 
could offer fewer galaxies as tracers of filaments, leading to incorrect determination of the orientation to adopt in the stack. In these cases, the geometric dilution will increase due to the inclusion of off-axis filaments in the stack. Nevertheless, in light of the projections above, we regard stacking as the most viable option to expand our search of the diffuse cosmic web. Alternatively, although observationally expensive, new ultra-deep observations ($\gg 50-60$ hr) targeting known overdensities would be fundamental to enlarge the sample of individual detections. Such efforts will remain essential before the advent of next-generation facilities, such as the BlueMUSE instrument currently under construction \citep{Richard2019} or the proposed WST \citep{bacon_wst_2024}. Complementary to direct imaging and stacking, \lya intensity mapping experiments offer another avenue to statistically constrain the properties of the IGM without resolving individual filaments \citep[e.g.,][]{LujanNiemeyer2026}.

\section{Summary and conclusions} \label{sect:conclusions}

We investigated the spatial distribution and physical properties of diffuse Ly$\alpha$ emission associated with cosmic filaments as traced by overdensities of LAEs at redshifts $3 \lesssim  z \lesssim  5$. The primary aim of this work was to move beyond extreme, quasar-dominated regions or protoclusters to systematically characterize the more typical cosmic web filaments that regulate galaxy assembly. 

We exploited the combined dataset from the MUDF and the MXDF, totaling $\gtrsim 250$~hours of ultra-deep integral-field spectroscopy and enabling the search of emission as low as $\approx 3-5\times 10^{-20}$~\sblcgs. To identify emitting structures, we first selected LAE overdensities with a friends-of-friends algorithm and then conducted a targeted search of low SB emission using the SHINE extraction code. 

The main findings of the analysis are as follows:
\begin{itemize}
\item We identified 41 significant LAE overdensities, probing cosmic environments  $\delta \approx 2-10$ times denser than the average field. Extended diffuse Ly$\alpha$ emission was detected in $\approx 74 \%$ of these overdensities, with the gas outside the CGM generally tracing the preferential filamentary axis defined by the positions of the member galaxies.
\item The observed SB varies systematically with different overdensities. Systems with $\delta \lesssim 2$ are largely non-detected at the depth of our observations, a large $\approx 1$~dex scatter in the detected SB is found for overdensities with $2 \lesssim \delta \lesssim 3.5$, while $\delta \gtrsim 3.5$ filaments show evident emission that plateaus at a maximum SB of $\approx 2\times 10^{-19}~$\sblcgs. Moreover, the total Ly$\alpha$ luminosity increases in denser regions, indicating that the emission in richer galaxy environments extends over larger areas.
\item The covering factor of emitting gas at the depth of our observation is $f_{\rm s}\approx 0.20-0.25$ within filaments. In a cosmological context, this covering factor translates to an incidence of \lya emitting regions of $\ell(\rm Ly\alpha)\approx 1.22$, which is consistent with the measured incidence of LLSs in absorption at comparable redshifts. 
\item Oriented stacking of filaments in $\delta \gtrsim 3.5$ overdensity revealed a coherent excess of signal, albeit diluted by the corresponding covering factor $f_{\rm s}\approx 0.20-0.25$. Based on this result, we computed a forecast for the detectability of cosmic web emission when stacking a larger sample of shallower MUSE observations (e.g., $\gtrsim 200-250$ for $4$ hr depth).
\end{itemize}

Through a systematic analysis of this complete and homogeneous search of extended \lya in the deepest available volumes observed by MUSE, we further our understanding of the empirical and physical properties of the cosmic web. Based on our findings, we corroborate the idea that LAE overdensities are good tracers of elongated filaments \citep{Bacon2021}. Comparative analysis of the observed SB in various overdensities suggests that the emitting gas is likely at the transition between optically thin and optically thick, maximizing the contribution of recombinations and collisional excitation while not strongly depending on the intensity of the local radiation field.  
This suggestion is further corroborated by the consistency of the incidence of emitting gas and LLSs. 

In light of these results, we conclude that the emerging picture of the cosmic web identifies these diffuse emitting regions as the denser spines of intergalactic filaments. These structures consist of gas that is partially ionized at moderate densities ($n_{\rm H} \approx 10^{-3}-10^{-1}$~cm$^{-3}$), with a likely additional contribution of embedded substructures.  This picture will have to be tested more quantitatively with the aid of detailed, cosmology-oriented radiative transfer calculations.

Future prospects for exploring the cosmic web rely heavily on expanding these samples through new observations. 
Since, in the near term, it is hard to imagine a substantial effort to increase the volume probed by ultra-deep data, we propose stacking as the most viable near-term strategy for exploiting existing medium-depth observations in the MUSE archive. Although this effort will be non trivial, our forecasts indicate the possibility of detecting the typical diffuse cosmic web over larger volumes. In the long term, the advent of next-generation facilities such as BlueMUSE or the WST will be essential for moving from statistical stacks to individual, three-dimensional maps of the intergalactic medium across an expanded volume.

\section*{Data availability}
The MUSE observations used in this work are publicly available in the ESO Science Archive (\url{https://archive.eso.org}) under program IDs 1100.A-0528 (MUDF), 1101.A-0127 (MXDF), and 094.A-0289, 095.A-0010, and 096.A-0045 (UDF-10 and MOSAIC). Reduced, science-grade data cubes are also available as Phase 3 data releases through the ESO Science Portal (\url{https://archive.eso.org/scienceportal}). The online material, consisting of the figures showing the \lya surface-brightness maps and spectra of all the LAE overdensities, is available on Zenodo at \url{https://doi.org/10.5281/zenodo.23104074}.

\begin{acknowledgements}
We thank the anonymous referee for the valuable comments that helped improve the manuscript. This work is supported by the Italian Ministry for Research and University (MUR) under Grant `Progetto Dipartimenti di Eccellenza 2023-2027' (BiCoQ). 
\end{acknowledgements}

\bibliographystyle{aa}
\bibliography{aa}

\begin{appendix}

\section{Properties of overdensities}\label{apx:ov-properties}

Table~\ref{tab:combined_overdensities} summarizes relevant properties for the detected overdensities in the MUDF and MXDF. 

\begin{table*}[h!]
\centering \small
\begin{threeparttable}
\caption{Combined sample of overdensities (MUDF+MXDF)}
\begin{tabular}{c c c c c c c c c c c}
\toprule
ID & $z$ & $\delta$ & $\sigma_{v}$ & $A_{\mathrm{tot}}$ & $A_{\mathrm{diff}}$ & $\mathrm{SB}_{\mathrm{int}}^\mathrm{diff}$ & $\log_{10}(L_{\mathrm{tot}})$ & Alignment & Sample & Note\\
 &  &  & $\rm km\,s^{-1}$ & $[\mathrm{arcsec}^2]$ & $[\mathrm{arcsec}^2]$ & $[10^{-18}\,\mathrm{erg\,s^{-1}\,cm^{-2}\,arcsec^{-2}}]$ & $[L_{\mathrm{tot}}\ \mathrm{in\ erg\,s^{-1}}]$ &  &  & \\
\midrule
$1$ & 2.813 & 2.4 & 192 & 548 & 311 & 0.19 $\pm$ 0.02 & 42.70 $\pm$ 0.05 & 0.65 & MUDF & -- \\
$2$ & 2.857 & 3.0 & 59 & 392 & 206 & 0.14 $\pm$ 0.02 & 42.37 $\pm$ 0.07 & 0.24 & MUDF & -- \\
$3$ & 2.996 & 3.3 & 205 & 598 & 293 & 0.02 $\pm$ 0.01 & 41.74 $\pm$ 0.10 & 0.70 & MXDF & -- \\
$4^*$ & 3.022 & 1.3 & 150 & 310 & 206 & 0.07 $\pm$ 0.01 & 42.04 $\pm$ 0.05 & 0.09 & MXDF & -- \\
$5$ & 3.047 & 4.3 & 241 & 1172 & 818 & 0.20 $\pm$ 0.01 & 43.11 $\pm$ 0.03 & 0.48 & MUDF & -- \\
$6$ & 3.067 & 1.7 & 213 & 594 & 409 & 0.08 $\pm$ 0.01 & 42.43 $\pm$ 0.04 & 0.67 & MXDF & -- \\
$7$ & 3.068 & 3.8 & 259 & 761 & 438 & 0.15 $\pm$ 0.02 & 42.73 $\pm$ 0.05 & 0.50 & MUDF & -- \\
$8$ & 3.085 & 1.6 & 172 & 124 & 60 & 0.03 $\pm$ 0.01 & 41.12 $\pm$ 0.21 & 0.55 & MXDF & -- \\
$9$ & 3.101 & 2.4 & 273 & 923 & 650 & 0.10 $\pm$ 0.01 & 42.69 $\pm$ 0.05 & 0.71 & MUDF & -- \\
$10$ & 3.125 & 3.5 & 84 & 804 & 494 & 0.23 $\pm$ 0.02 & 42.95 $\pm$ 0.03 & 0.82 & MUDF & -- \\
$11$ & 3.172 & 1.6 & 265 & 160 & 58 & 0.05 $\pm$ 0.01 & 41.39 $\pm$ 0.10 & 0.75 & MXDF & -- \\
$12$ & 3.190 & 3.2 & 137 & 363 & 137 & 0.04 $\pm$ 0.01 & 41.66 $\pm$ 0.14 & 0.25 & MXDF & -- \\
$13$ & 3.193 & 1.0 & 81 & 219 & 132 & 0.04 $\pm$ 0.01 & 41.63 $\pm$ 0.11 & 0.98 & MUDF & -- \\
$14$ & 3.233 & 4.1 & 277 & 3044 & 1507 & 0.17 $\pm$ 0.01 & 43.30 $\pm$ 0.02 & 0.38 & MUDF & -- \\
$15$ & 3.262 & 1.8 & 74 & 245 & 95 & 0.30 $\pm$ 0.05 & 42.35 $\pm$ 0.07 & 0.76 & MUDF & -- \\
$16$ & 3.332 & 2.2 & 392 & 321 & 244 & 0.07 $\pm$ 0.01 & 42.13 $\pm$ 0.07 & 0.36 & MXDF & -- \\
$17$ & 3.348 & 2.4 & 33 & 829 & 550 & 0.14 $\pm$ 0.02 & 42.76 $\pm$ 0.05 & 0.68 & MUDF & -- \\
$18$ & 3.416 & 1.3 & 105 & -- & -- & -- & -- & 0.49 & MXDF & -- \\
$19$ & 3.417 & 1.5 & 94 & 372 & 181 & 0.04 $\pm$ 0.02 & 41.74 $\pm$ 0.16 & 0.51 & MUDF & -- \\
$20$ & 3.433 & 3.1 & 288 & 328 & 115 & 0.06 $\pm$ 0.02 & 41.70 $\pm$ 0.16 & 0.57 & MXDF & -- \\
$21$ & 3.466 & 3.3 & 402 & 273 & 89 & 0.07 $\pm$ 0.01 & 41.69 $\pm$ 0.08 & 0.69 & MXDF & -- \\
$22$ & 3.540 & 1.0 & 16 & -- & -- & -- & -- & 0.79 & MUDF & -- \\
$23$ & 3.558 & 2.1 & 305 & 196 & 59 & 0.08 $\pm$ 0.02 & 41.55 $\pm$ 0.11 & 0.24 & MXDF & -- \\
$24$ & 3.603 & 4.3 & 346 & -- & -- & -- & -- & 0.26 & MXDF & Sky \\
$25$ & 3.669 & 3.9 & 388 & -- & -- & -- & -- & 0.24 & MXDF & -- \\
$26$ & 3.714 & 5.8 & 637 & 654 & 190 & 0.28 $\pm$ 0.01 & 42.57 $\pm$ 0.02 & 0.56 & MXDF & -- \\
$27^*$ & 3.763 & 1.2 & 200 & 242 & 130 & 0.33 $\pm$ 0.01 & 42.47 $\pm$ 0.02 & 0.08 & MXDF & -- \\
$28$ & 4.007 & 8.0 & 243 & 2149 & 1282 & 0.18 $\pm$ 0.01 & 43.18 $\pm$ 0.03 & 0.52 & MUDF & -- \\
$29$ & 4.041 & 1.0 & 106 & -- & -- & -- & -- & 0.79 & MUDF & -- \\
$30$ & 4.048 & 1.6 & 207 & -- & -- & -- & -- & 0.65 & MXDF & -- \\
$31^*$ & 4.137 & 3.2 & 132 & -- & -- & -- & -- & 0.42 & MXDF & -- \\
$32^*$ & 4.155 & 1.0 & 229 & -- & -- & -- & -- & 0.73 & MXDF & -- \\
$33$ & 4.279 & 2.2 & 227 & 271 & 140 & 0.25 $\pm$ 0.03 & 42.34 $\pm$ 0.05 & 0.48 & MXDF & -- \\
$34^*$ & 4.369 & 1.3 & 160 & -- & -- & -- & -- & 0.62 & MXDF & -- \\
$35^*$ & 4.408 & 1.5 & 156 & -- & -- & -- & -- & 0.45 & MXDF & Sky \\
$36$ & 4.422 & 2.8 & 129 & -- & -- & -- & -- & 0.71 & MUDF & Sky \\
$37$ & 4.444 & 4.8 & 91 & -- & -- & -- & -- & 0.45 & MUDF & Sky \\
$38$ & 4.471 & 1.4 & 44 & -- & -- & -- & -- & 0.44 & MXDF & -- \\
$39$ & 4.516 & 4.0 & 456 & -- & -- & -- & -- & 0.09 & MXDF & Sky \\
$40^*$ & 4.761 & 2.1 & 828 & -- & -- & -- & -- & 0.27 & MXDF & Sky \\
$41$ & 4.940 & 1.4 & 182 & 340 & 193 & 0.12 $\pm$ 0.02 & 42.09 $\pm$ 0.07 & 0.87 & MXDF & -- \\
\bottomrule
\label{tab:combined_overdensities}
\end{tabular}
\begin{tablenotes}
    \small
    \item Notes. ID: group identification number (asterisks denote additional groups relative to \citealp{Bacon2021}); $z$: spectroscopic redshift; $\delta$: overdensity parameter; $\sigma_v$: group velocity dispersion; $A_{\mathrm{tot}}$: total emission area; $A_{\mathrm{diff}}$: area of the diffuse component; $\mathrm{SB}_{\mathrm{int}}^{\mathrm{diff}}$: average intrinsic SB of the diffuse component; $\log_{10}(L_{\mathrm{tot}})$: luminosity of the diffuse component; Alignment: overdensity alignment factor; Sample: survey subsample (MUDF or MXDF); Note: additional flags.
\end{tablenotes}
\end{threeparttable}
\end{table*}

\section{Signal extraction using a wavelet analysis}\label{apx:wavelet}
To independently verify the presence of extended \lya emission in our MUDF sample, we perform a multiscale analysis using the Isotropic Undecimated Wavelet Transform (IUWT), following the methodology described in \cite{Bacon2021}. This approach is specifically designed to disentangle diffuse, low SB signals from the overlap of compact bright sources (e.g., LAEs and foreground and background objects). 

The IUWT decomposes an NB image into coefficient levels, each capturing structures at different spatial scales. We apply this filtering to an NB S/N image constructed by collapsing a spectral window of $8$~cMpc centered on the mean redshift of each overdensity. The spatial structures are filtered into three primary bands: (i) high-frequency signal corresponding to noise, in the range $\approx 0.2 - 0.6$ arcsec; (ii) medium-frequency signal, in the range $0.6-2.2$ arcsec, which captures emission from compact sources such as individual LAEs; (iii) Low-frequency (diffuse signal): structures in the range $2.2-37$ arcsec, optimized fro the detection of extended emission. 

The extraction pipeline consists of the following steps: firstly, we apply the IUWT to the original S/N image, and we perform a segmentation of the medium-frequency to identify all compact sources. Secondly, we use the resulting segmentation map to mask the compact signal in the original S/N image by replacing the values with a local average S/N estimate. Finally, we apply the IUWT again to the masked S/N image and use the low-frequency to define the boundaries of the extended \lya emission (see \citealt{Bacon2021} for all the details).

\begin{figure*}
    \sidecaption
    \begin{minipage}[b]{10cm}
    \subfloat[]{\includegraphics[width=10cm]{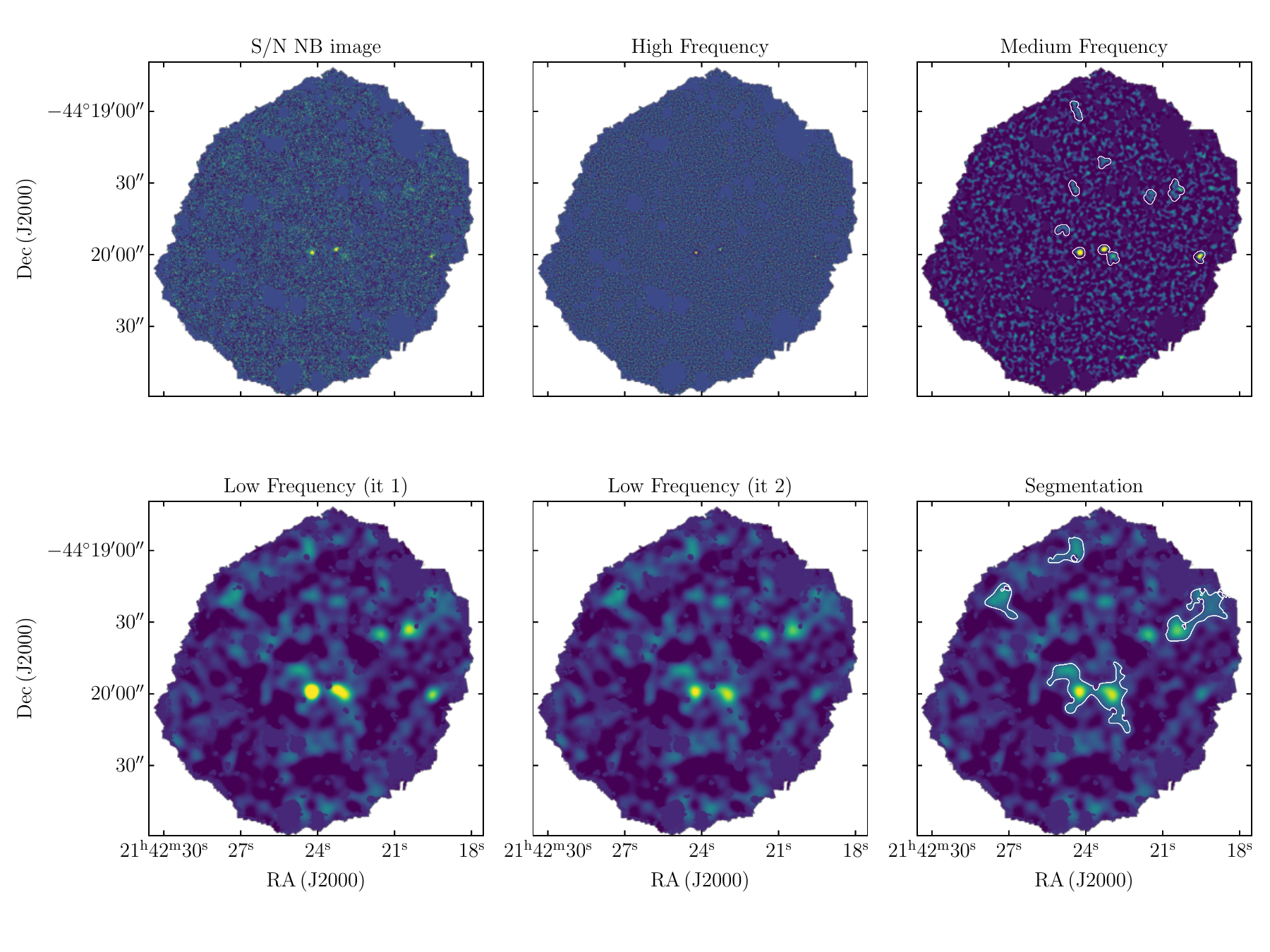}\label{fig:wl3.05}}\\
    \subfloat[]{\includegraphics[width=10cm]{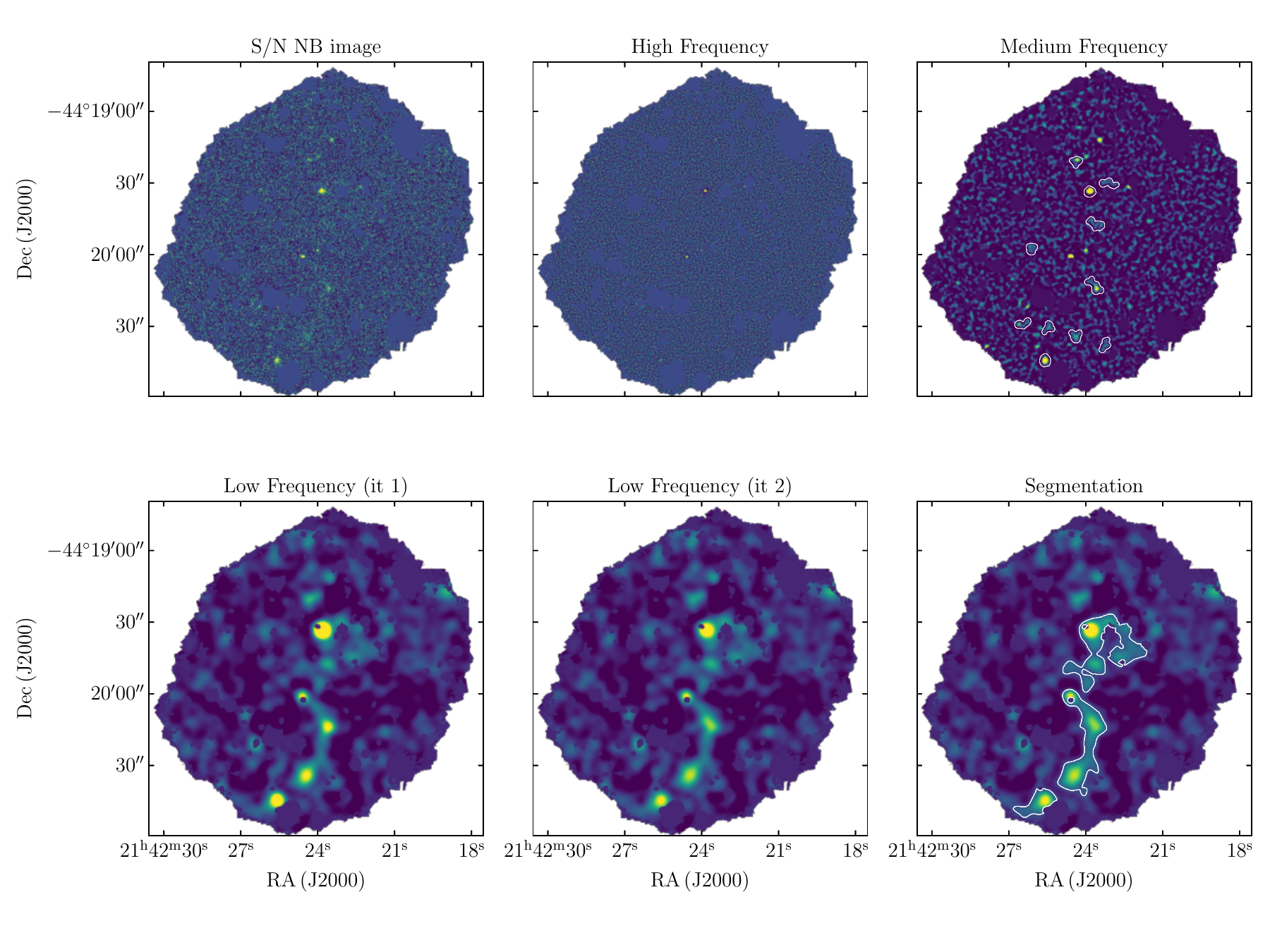}\label{fig:wl2.81}}
    \end{minipage}
    \caption{Results of the wavelet-based IUWT analysis detailed in \cite{Bacon2021} for two representative LAE overdensities in the MUDF at $z=2.813$ and $z=3.047$. Panels (a) and (b) show the narrow-band S/N images after multiscale decomposition, with the low-frequency component highlighting the diffuse \lya emission after masking compact sources. The extracted extended emission is shown as detected by the IUWT filtering, enabling a direct comparison with the SHINE-based voxel extraction discussed in the text.}
    \label{fig:wavelet-analysis}
\end{figure*}

In Fig.~\ref{fig:wavelet-analysis}, we compare the extended emission maps obtained with SHINE against those produced by the wavelet-based IUWT analysis. We find general consistency in the detection of the primary, high-S/N-emitting structures. The same level of agreement is observed for the filamentary structure extracted at $z=3.067$ in the MXDF, which is discussed in detail in \citet{Bacon2021} and named GID 2 (see their Figure 12). However, some morphological differences exist, which can be attributed to the inherent algorithmic differences between voxel-based grouping (SHINE) and multiscale filtering (IUWT) approaches. SHINE is a three-dimensional grouping algorithm that identifies connected structures by leveraging the full spectral information across the data cube (typically $\gtrsim 25$ channels for each overdensity). In contrast, IUWT operates on a two-dimensional S/N image collapsed over a specific velocity window (e.g., $8$ cMpc, corresponding to $8-9$ channels at $3\lesssim z \lesssim 4$). This can explain why, in some cases, SHINE recovers emissions missed by IUWT. 

Moreover, the two codes approach the signal's scale differently. IUWT intrinsically isolates extended, diffuse, and low-SB signal within the low-frequency band (on scales $\gtrsim 15$ arcsec). This results in smoother, more continuous structures. SHINE grows structures by grouping nearby voxels from the individual voxel scale ($0.2$ arcsec) to larger scales ($\gtrsim 15$ arcsec). Consequently, SHINE often maps less-smoothed structures and may lack some of the diffuse fill-in captured by wavelet filtering. Finally, SHINE uses the local variance of each individual voxel, naturally taking into account the varying exposure time. This is particularly advantageous in fields with a non uniform depth, such as the MUDF. IUWT estimates a global threshold based on the standard deviation computed from the entire wavelet-band image and is therefore best suited for nearly uniform-depth observation. In regions with strongly varying exposure times, a global threshold may not perfectly track local noise fluctuations. This can lead to different behaviors at the field edges: IUWT may, in some cases, pick up low significance features if the global threshold is influenced by the higher sensitivity of the central regions. 

All considered, the agreement between the two methods on the main features provides a robust benchmark for our results. In the analysis, we therefore rely on our SHINE extraction, which has also been validated by visual inspection of the spectra.

\section{Overdensity stacking details}\label{apx:stack-details}

In Table \ref{tab:stack_info}, we provide the detailed parameters used for the oriented stacking analysis. 

\begin{table*}
\centering 
\caption{Properties of the high-overdensity subsample ($\delta \gtrsim 3.5$) used for the oriented stacking analysis.}
\label{tab:stack_info}
\begin{tabular}{llcccc}
\toprule
ID & $\delta$ & Dataset & RA [deg] & Dec [deg] & Angle [deg] \\
\midrule
5 & 4.3 & MUDF & 325.601282 & -44.334963 & 80.8 \\
7 & 3.8 & MUDF & 325.607695 & -44.331862 & 86.8 \\
10 & 3.5 & MUDF & 325.603827 & -44.330613 & -70.4 \\
14 & 4.1 & MUDF & 325.599819 & -44.331374 & 50.4 \\
25 & 3.9 & MXDF & 53.171908 & -27.782074 & 18.9 \\
26 & 5.8 & MXDF & 53.165507 & -27.783930 & 68.3 \\
28 & 8.0 & MUDF & 325.604187 & -44.334783 & 66.4 \\
\bottomrule
\end{tabular}
\begin{tablenotes}
    \small
    \item Notes. ID: group identification number; $\delta$: overdensity parameter; RA and Dec: coordinates of the centroids; Angle: rotation angle relative to the horizontal axis.
\end{tablenotes}
\end{table*}

\section{Selection functions of UDF-10 and MOSAIC}\label{apx:selfunc}

\begin{figure*}
\sidecaption
\includegraphics[width=12cm]{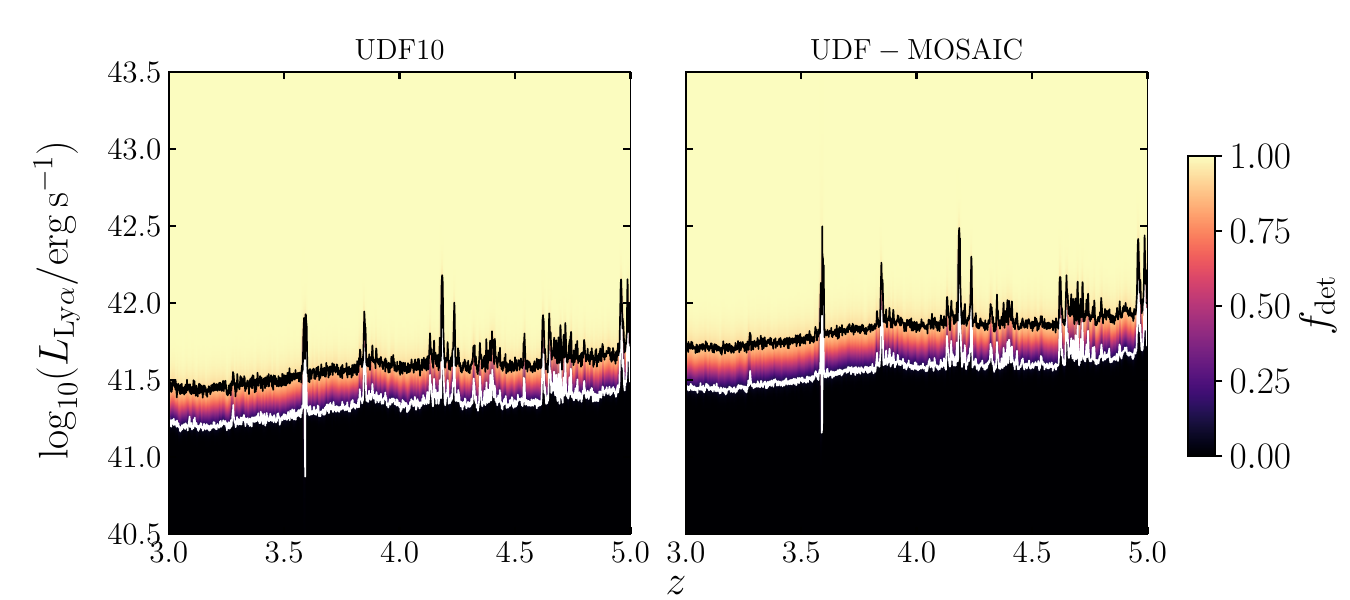}
\caption{Selection functions derived for LAEs analyzed in the UDF-10 and MOSAIC regions within the selected area shown in Fig. \ref{fig:mxdf-footprint}. The white and black contours indicate the 10\% and 90\% completeness limits, respectively.}
\label{fig:selfunc}
\end{figure*}

\section{Emission maps of the full sample of overdensities}\label{apx:gallery}

\begin{figure*}
\centering
\includegraphics[width=\hsize]{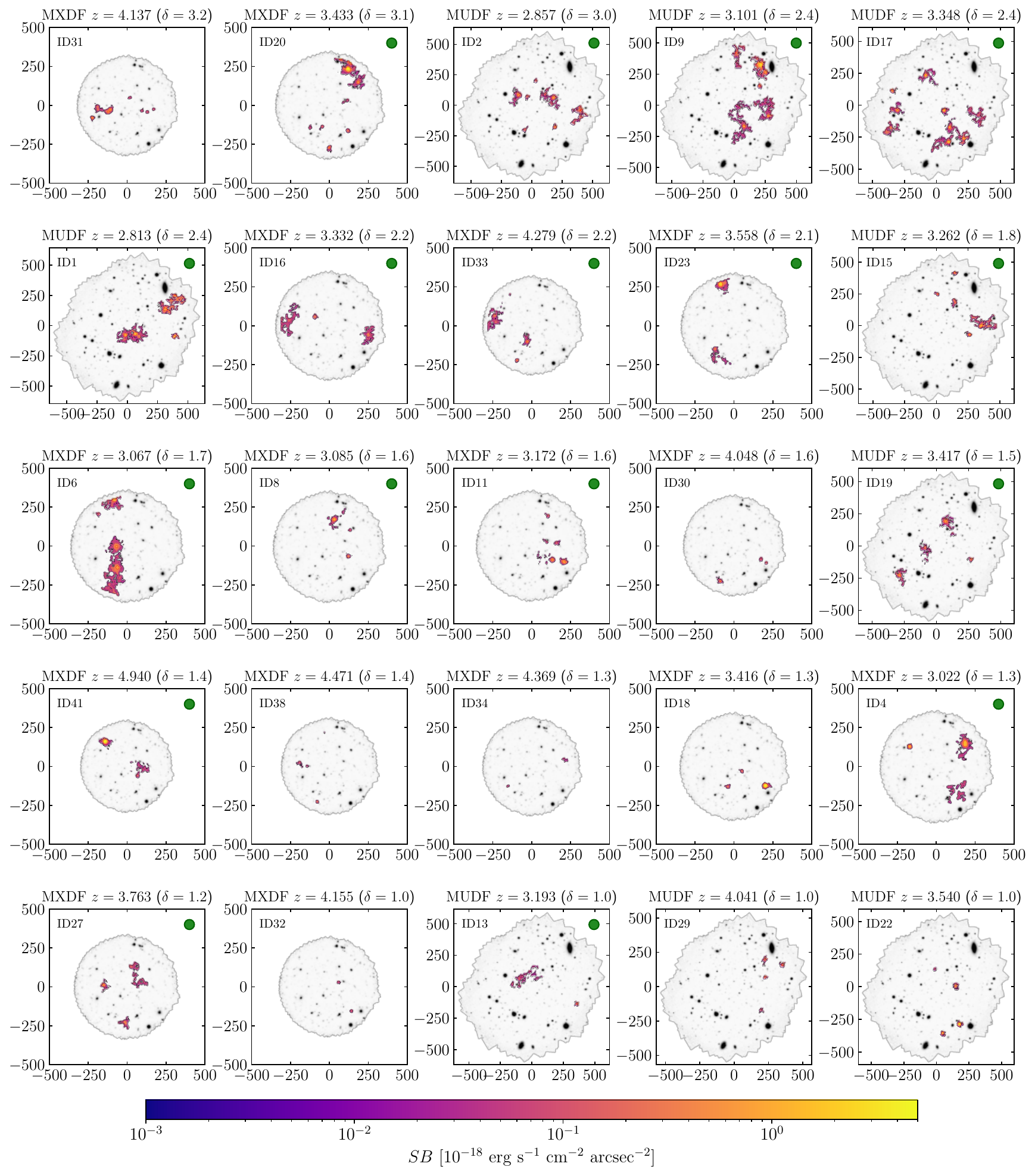}
\caption{Same as Fig.~\ref{fig:gallery}, but for the remaining LAE overdensities of the sample, sorted by decreasing $\delta_\mathrm{LAE}$.}
\label{fig:gallery-full}
\end{figure*}

In Fig.~\ref{fig:gallery-full}, we present the \lya SB maps of the overdensities not shown in Fig.~\ref{fig:gallery}, continuing the sequence sorted by decreasing $\delta_\mathrm{LAE}$.

\end{appendix}

\end{document}